\documentclass[12pt,preprint]{aastex}

\def \etal{{et~al.\null}}

\def\lea{\mathrel{<\kern-1.0em\lower0.9ex\hbox{$\sim$}}}
\def\gea{\mathrel{>\kern-1.0em\lower0.9ex\hbox{$\sim$}}}
\newcommand{\lta}{{\>\rlap{\raise2pt\hbox{$<$}}\lower3pt\hbox{$\sim$}\>}}
\newcommand{\gta}{{\>\rlap{\raise2pt\hbox{$>$}}\lower3pt\hbox{$\sim$}\>}}

\shorttitle{Using H\boldmath{$\alpha$} Morphology and Surface Brightness Fluctuations}
\author{
Bradley~C.~Whitmore,\altaffilmark{1} 
Rupali~Chandar,\altaffilmark{2} 
Hwihyun~Kim,\altaffilmark{3} 
Catherine~Kaleida,\altaffilmark{3} 
Max~Mutchler,\altaffilmark{1}
Daniela~Calzetti,\altaffilmark{4} 
Abhijit~Saha,\altaffilmark{5}
Robert~O'Connell,\altaffilmark{6}
Bruce~Balick,\altaffilmark{7}
Howard~E.~Bond,\altaffilmark{1}
Marcella~Carollo,\altaffilmark{8}
Michael~J.~Disney,\altaffilmark{9}
Michael~A.~Dopita,\altaffilmark{10}
Jay~A.~Frogel,\altaffilmark{11}
Donald~N.~B.~Hall,\altaffilmark{12}
Jon~A.~Holtzman,\altaffilmark{13}
Randy~A.~Kimble,\altaffilmark{14}
Patrick~J.~McCarthy,\altaffilmark{15}
Francesco~Paresce,\altaffilmark{16}
Joseph~I.~Silk,\altaffilmark{17}
John~T.~Trauger,\altaffilmark{18}
Alistair~R.~Walker,\altaffilmark{19}
Rogier~A.~Windhorst,\altaffilmark{3} and
Erick~T.~Young\/\altaffilmark{20}}

\shortauthors{Whitmore et al.}
\email{whitmore@stsci.edu}
\altaffiltext{1}{Space Telescope Science Institute, Baltimore, MD, USA}
\altaffiltext{2}{Department of Physics \& Astronomy, The University of Toledo, Toledo, OH 43606}
\altaffiltext{3}{School of Earth and Space Exploration, Arizona State University, Tempe, AZ 85287-1404, USA}
\altaffiltext{4}{Department of Astronomy, University of Massachusetts, Amherst, MA 01003, USA}
\altaffiltext{5}{NOAO, Tucson, AZ 85726-6732, USA}
\altaffiltext{6}{Department of Astronomy, University of Virginia,  Charlottesville, VA 22904-4325, USA}
\altaffiltext{7}{Department of Astronomy, University of Washington,  Seattle, WA 98195-1580, USA}
\altaffiltext{8}{Institute of Astronomy, ETH-Zurich, Zurich, 8093  Switzerland}
\altaffiltext{9}{Department of Physics and Astronomy, Cardiff University,  Cardiff CF24 3AA, UK}
\altaffiltext{10}{Research School of Astronomy \& Astrophysics, The  Australian National University, Cotter Road, Weston Creek, ACT  2611, Australia}
\altaffiltext{11}{AURA, Washington, DC 20005, USA}
\altaffiltext{12}{Institute for Astronomy, Honolulu, HI 96822, USA}
\altaffiltext{13}{New Mexico State University, Las Cruces, NM 88003,  USA}
\altaffiltext{14}{Goddard Space Flight Center, Greenbelt, MD 20771, USA}
\altaffiltext{15}{Carnegie Institute of Washington, Pasadena, CA  91101-1292, USA}
\altaffiltext{16}{Istituto di Astrofisica Spaziale e Fisica Cosmica, INAF, Via Gobetti 101, 40129, Bologna, Italy}
\altaffiltext{17}{Department of Physics, University of Oxford, Oxford  OX1 3PU, UK}
\altaffiltext{18}{NASA JPL, Pasadena, CA 91109, USA}
\altaffiltext{19}{Cerro Tololo Inter-American Observatory,
La Serena, Chile}
\altaffiltext{20}{NASA--Ames Research Center, Moffett Field, CA 94035, USA}

\begin{document}


\title{Using H$\alpha$ Morphology and Surface Brightness Fluctuations to Age-Date Star Clusters in M83}

\begin{abstract}

We use new WFC3 observations of the nearby grand design spiral galaxy
M83 to develop two independent methods for estimating the ages of
young star clusters.  The first method uses the physical extent and
morphology of H$\alpha$ emission to estimate the ages of clusters
younger than $\tau \approx10$~Myr.  It is based on the simple premise
that the gas in very young ($\tau <$~few Myr) clusters is largely
coincident with the cluster stars, is in a small, ring-like structure
surrounding the stars in slightly older clusters since massive star
winds and supernovae have had time to push out the natal gas (e.g.,
$\tau \approx5$~Myr), and is in a larger ring-like bubble for still
older clusters (i.e., $\approx5$--10 Myr). If no H$\alpha$ is
associated with a cluster it is older than $\approx10$ Myr.  The
second method is based on an observed relation between pixel-to-pixel
flux variations within clusters and their ages.  This method relies on
the fact that the brightest individual stars in a cluster are most
prominent at ages around 10 Myr, and fall below the detection limit
(i.e., M$_V < -3.5$) for ages older than about 100 Myr.  Older
clusters therefore have a smoother appearance and smaller
pixel-to-pixel variations. The youngest clusters also have lower flux
variations, hence the relationship is double valued.  This degeneracy
in age can be broken using other age indicators such as H$\alpha$
morphology.  These two methods are the basis for a new morphological
classification system which can be used to estimate the ages of star
clusters based on their appearance.  We compare
previous age estimates of clusters in M83 determined from fitting
\textit{UBVI}H$\alpha$ measurements using predictions from stellar evolutionary
models with our new morphological categories and find good agreement
at the $\approx95$\% level. The scatter within categories is
$\approx0.1$~dex in $\log\tau$ for young clusters ($<$10 Myr) and
$\approx0.5$~dex for older ($>$10 Myr) clusters.  A by-product of
this study is the identification of 22 ``single-star'' HII regions in
M83, with central stars having ages $\approx4$ Myr.

\end{abstract}


\section{Introduction}

Age estimates are required for studying the evolutionary history of 
star clusters. Based on such age estimates,
a general framework has been developed
starting from formation within the dense cores of
giant molecular clouds (GMCs); 
a stage where the young stars are completely
obscured by their dusty gas cocoons; 
an emerging stage where the clusters
become visible in the infrared (IR) and then visible in the optical as stellar winds and supernovae blow away the gas and dust; a stage where an expanding 
bubble of ionized gas is blown; and later stages with no evidence of
ionized gas (e.g.,
see the Lada \& Lada 2003 review article). 

In the past, two general
methods for estimating the ages of unresolved extragalactic star clusters have
been used.  The first requires obtaining high quality spectroscopic
observations covering wavelength regions of lines which change as a
function of time.  Examples of this approach include Bica \& Alloin (1986), Schweizer \&
Seitzer (1993), Whitmore \etal\/ (1999), Bastian et~al.\  (2009) and
Wofford et~al.\ (2010). While these spectroscopic observations
typically provide high quality age determinations, they are limited to
age-dating relatively small numbers of bright clusters due to
constraints on 
the brightness of clusters that can be observed spectroscopically in a
reasonable amount of time. The
standard method for age-dating large numbers of extra-galactic
clusters compares photometry in several broadband filters (e.g., \textit{UBVI})
with predictions from population synthesis models (e.g., Chandar
et~al.\ 2010).

Comparisons of age estimates using spectroscopic and photometric 
observations of the same clusters show
good agreement in most cases. 
An early example of this is the famous Searle et~al.\ (1980)
paper where they compare integrated four-filter \textit{ugvr} photometry
of 61 star clusters in the Magellanic Clouds with the strength of different
spectral features (e.g., Balmer lines, $G$~band, etc.)
A recent example is Wofford et~al.\ (2010) who compared 
ages estimated from spectral energy distribution (SED) fitting with those derived
from UV spectroscopy for 14 young ($\tau \lea 30$~Myr) clusters
in the nuclear starburst region of M83
and found that the photometric ages
are within a factor of 1.4 of the spectroscopic
ones. Whitmore et~al.\ (2010) find similar agreement between their
photometric age estimates and the spectroscopic
estimates from Bastian et~al.\ (2009) for clusters in the Antennae galaxies.
While generally satisfactory, both the spectroscopic and
photometric techniques have limitations.
It is therefore important to develop
independent methods for estimating the ages of clusters, especially for
cases where spectroscopic or multi-band photometric observations are not
available and it is necessary to estimate ages based on morphological
appearance alone.

Here, we develop two new methods for estimating the ages of
young star clusters in nearby galaxies based on high-resolution
images at optical wavelengths.  The first method uses the morphology
of the ionized gas and its position relative to the cluster stars, as measured
from narrowband H$\alpha$ emission, to estimate ages ($\tau$) for clusters
younger than $\tau \lea 10$~Myr. 
This method relies on the general premise that the distribution of
H$\alpha$ will be  largely coincident with the distribution of optical light
in the youngest clusters (i.e., $<$~few Myr), will be in a small ring-like
structure around the optical stellar emission in slightly older clusters where
massive star winds and supernovae have had time to blow a bubble (i.e., $5 \approx$ Myr), and will be in a larger ring-like bubble for still older clusters
(i.e., 5--10 Myr).

Many past observational and theoretical studies of 
 HII regions and ``supershells''
in the Milky Way and nearby galaxies have laid the
groundwork for this method. 
For example, Walborn (2002; see also Walborn \& Parker 1992 and especially
Walborn 2010 for related discussions) 
outlined an evolutionary cluster sequence 
based on observed properties of
several well known clusters and OB associations in the Milky Way
with ages ranging from $\approx1$~Myr to $\approx10$~Myr. This sequence
was then used  
to illustrate observed changes with age in the visually brightest stars,
in the ionized gas and dust content, and in the existence
of red supergiants. 
Much of this sequence was based on spectroscopy of individual
stars. While we are unable to make similarly detailed observations
at the distance of M83, many of the same basic correlations and
underlying physical processes are relevant for the age sequence
outlined in this paper.
Theoretically, several works have made predictions for the
size evolution of an expanding HII region over time.
Oey \& Clark (1997, 1998) model the size evolution
due to mass loss and supernova-injected energy from cluster stars,
and assume that expanding bubbles 
``stall'' when their internal pressure equals the ambient pressure
in the ISM. 
These simulations predict a strong dependence of bubble
size on both the age and mass of the central cluster.
(e.g., Weaver et~al.\ 1977, Oey \& Massey 1995, Oey \& Garcia-Segura
2004, Dopita et~al.\ 2006a, 2006b).
Our observations of clusters in M83 reveal a strong dependence of
bubble size on cluster age, and possibly a weak dependence on 
cluster mass, as discussed in Section~6.

The second 
method developed in this paper uses the surface brightness fluctuations of cluster stars to estimate their ages.   Young  clusters have
strong pixel-to-pixel flux variations, due to the presence of 
massive, luminous stars.  As a cluster ages, the bright, short-lived massive stars 
disappear, and these fluctuations fade in strength.
This technique is especially useful
in the range $\tau\approx 10$--100~Myr,
ages that can be somewhat difficult to deal with 
using the SED method, because the predicted integrated
colors loop back on themselves.

Our primary target for this study is the spiral galaxy M83.
At a distance of 4.5 Mpc (corresponding to a distance modulus of 
$m-M = 28.28$; Thim et~al.\ 2003, 
and a pixel scale of 0.876 pc~pixel$^{-1}$), M83, nicknamed the
``Southern Pinwheel,'' is the nearest massive grand-design spiral
galaxy. It is a slightly barred galaxy, with a Hubble type SAB(s)c (3RC).
In this work, we make use of observations taken with the Wide-Field
Camera 3 (WFC3), which were described in detail in Dopita et~al.\ (2010)
and Chandar et~al.\ (2010). Briefly, the observations are part of the
Early Release Science project~1 (ERS1) program 11360 (PI: O'Connell)
and were taken in August, 2009. 
Observations were taken in several broadband
(``UV'' - F225W, ``U'' - F336W, ``B'' - F438W, ``V'' - F555W, ``I'' -
F814W, ``J'' - F110W, and ``H'' - F160W), and narrowband filters
([OIII] - F373N, H$\beta$ - F487N, [OII] F502N, H$\alpha$ - F657N,
[SII] - F673N, Paschen$\beta$ - F128N, and [FeII] - F164N).
In the Appendix we briefly investigate clusters in M51 and find that    
the age versus morphological category relationship
derived for M83 is appropriate for this galaxy as well. 

This paper is organized as follows:
In Section~2 we outline our working scheme for an 
evolutionary cluster classification system based
on observables at optical wavelengths. In Section~3 we investigate the
correlation between H$\alpha$ 
morphology and SED age estimates while
Section~4 examines the correlation between the strength of surface brightness
fluctuations within clusters and their ages. 
A catalog of smaller HII regions, apparently ionized by  ``single
stars,'' is presented in Section~5.  
We examine the effect of cluster mass on 
H$\alpha$ bubble size in Section~6, and summarize
our primary results in Section~7.

\section{An Evolutionary Classification Scheme Based on Observables at Optical Wavelengths}
The early  formative stages of star cluster evolution 
are best studied in the IR or microwave part of the spectrum, 
since the stars remain embedded in  their
placental dust cocoons for the first million years or so (Lada \& Lada 2003). 
For example, in nearby groups and clusters in the Milky Way, 
individual young stellar objects (YSOs) 
can be age-dated using a classification scheme ranging from  
category 0 to III, as developed by Wilking et~al.\ (1989) and 
Andre et~al.\ (1993), based on near-IR observations.

In this paper we outline a working classification scheme  for cluster evolution
in external galaxies. We focus 
on the later stages based on the optical portion of the spectrum.
Our primary goals are
to test how well we can use H$\alpha$ morphology and surface brightness fluctuations to age date
star clusters. The basic categories are defined below. In all cases we 
assume that the clusters have radial profiles that are broader than the point-spread function (PSF),
hence confusion with individual stars is not a major issue
(see Chandar et~al.\ 2010).  
Examples of clusters in categories~3 through 6 are shown in Figure~\ref{fig:18plot}.  M83 clusters in categories~1 and 2 will be 
discussed in a later paper which presents our WFC3 observations in the
$J$ and $H$~bands.


{\noindent}{\em Category 1:} {\bf opaque dust cloud} - core of a GMC -
dark region on optical image with  no associated IR/UVIS source - often studied using millimeter 
observations of the HCN line (e.g., Gao \& Solomon 2004)

\vskip 0.1in

{\noindent}{\em Category 2a:} - {\bf embedded cluster} - weak IR source with no optical (I band)
counterpart  (i.e., Av $>$ 10) - no ionized H$\alpha$ gas visible

\vskip 0.1in

{\noindent}{\em Category 2b:} - {\bf obscured cluster} - strong IR and weak optical source (3 $<$ Av $<$ 10)

\vskip 0.1in

{\noindent}{\em Category 3:} - {\bf emerging cluster} - ionized gas spatially coincident
with cluster stars - cluster has a reddish color due to dust - surface
brightness fluctuations from individual stars are relatively small
since the brightest, evolved stars have not yet appeared

\vskip 0.1in

{\noindent}{\em Category 4a:} - {\bf very young cluster} - ionized gas in a small bubble
surrounding the cluster  - cluster has a bluish color since most
of the dust has been expelled - surface brightness fluctuations from
individual stars are strong

\vskip 0.1in

{\noindent}{\em Category 4b:} - {\bf young cluster} - similar to category 4a but 
the ionized gas is now in a large bubble surrounding
the cluster (i.e., radii larger than approximately 20 pc) - surface brightness fluctuations from
individual stars reach maximum

\vskip 0.1in

{\noindent}{\em Category 5a:} - {\bf young/intermediate-age cluster}  - no ionized gas  is observed - 
surface brightness fluctuations are still present, but weaker

\vskip 0.1in

{\noindent}{\em Category 5b:} - {\bf intermediate-age cluster} -
ionized gas and surface brightness fluctuations
among cluster stars are not observed - cluster has a slightly redder color 
 due to aging of stars 

\vskip 0.1in

{\noindent}{\em Category 6:} - {\bf old cluster} - no ionized gas or
surface brightness fluctuations among cluster stars are observed,  - 
cluster appears yellow/red with no evidence of dust in vicinity

\section{H\boldmath{$\alpha$} Morphology as an Age Indicator for \boldmath{$\tau\lea 10$}~Myr Clusters}

\subsection{General Trends}
In principle, it should be possible to approximately estimate 
the ages of young ($\tau \lea 10$~Myr) star clusters
from the size of the ionized gas (H$\alpha$) bubble that surrounds
them, as outlined in Sections~1 and 2.
In this section we test this idea by selecting 
a representative catalog of young clusters primarily based on their apprearance 
in the narrowband H$\alpha$ image
(i.e., the red color in Figure~\ref{fig:large_image}), and
then visually classify each source based on the
scheme outlined in Section~2.
Each region is then matched with the apparent central source of ionization
from the Chandar et~al.\ (2010) catalog, and compared with the ages from that
paper based on SED fitting of the \textit{UBVI}H$\alpha$ bands.
We do not, at this point, separate the clusters by mass (the effect
of cluster mass on bubble size is discussed in Section~6).
Basic properties of the selected clusters are presented in Table~1,
including position (RA and decl.) and photometric
measurements ($M_V$, $V\!-\!I$, and $U\!-\!B$) for the central cluster.

The top row of Figure~\ref{fig:18plot} shows examples of clusters in
categories 3-6, with different H$\alpha$ morphologies
as outlined in Section~2.
The bottom set of panels shows the location of each cluster in the category
in a  $V\!-\!I$ versus $U\!-\!B$ two-color diagram, where the photometry is drawn
from the catalog described in Chandar et~al.\ (2010).
These panels show how well our morphological categories
group the clusters in color-color space. 
Predictions from  single stellar population
G.~Bruzual \& S.~Charlot (2009; private communications, hereafter BC09)
models are superposed for comparison. As expected, clusters in
the two youngest categories (3 and 4a)
have somewhat higher extinction than those
in later categories, which results in their colors
spreading out along the reddening vector.

The middle set of panels present histograms of ages determined
from the photometric dating method used in
Chandar et~al.\ (2010), for all clusters within a given category.
These age estimates are included in Table~1.
There is a clear trend for clusters in later categories
to have older ages.
For clusters with associated H$\alpha$ 
emission, i.e., categories 3, 4a, 4b,
we find an rms scatter of approximately 0.1 in $\log(\tau/\mbox{yr})$
within each category.
While our SED dating method includes photometry in the narrowband
filter directly in the fit, the photometry is measured
in a small (3~pixel radius) aperture that misses most or all of the H$\alpha$ 
emission after Category~3.
Hence the age estimates from the SED fitting and from
the H$\alpha$ morphology are largely independent
of one another. The rms scatter in photometric ages for
clusters identified within morphological categories 5 and 6 increases 
to $\approx0.5$ in $\log(\tau/\mbox{yr})$,
primarily because morphology becomes a cruder estimate in this age
regime.

The good overall correlation between morphological classification and
photometric age estimates from Chandar et~al.\ (2010) is also apparent in
Figure~\ref{fig:large_image}.
This figure includes the SED ages for the clusters where
morphological classifications have been 
determined  using the prescripts defined in Section~2. The region
covers a $1.0\times1.0$~kpc$^2$ area of M83. 
We note in particular the intermediate-age and old clusters
in the upper central part of the image which have a fuzzy appearance due to the absence of bright young stars. 
The Appendix shows similar figures for selected regions in M51.

Figure~\ref{fig:ha_vs_sed} plots the 
morphological category assigned here versus $\log(\tau/$yr)
based on the SED age estimates from Chandar et~al.\ (2010). 
The top panel shows the full range of cluster ages and categories.
There is a clear, albeit nonlinear, correlation,
with a steeper correlation between morphological category and
age for clusters younger than $\approx10^7$~yr.
The nonlinearity is to be   expected since H$\alpha$ morphology evolves 
rapidly between the ages of 1--10 Myr.
Note that in Figure~\ref{fig:ha_vs_sed} we have 
subdivided the morphological types into a finer grid of categories
(based on subjective estimates of the H$\alpha$ bubble size)
than outlined in Section~2. 
The bottom panel of Figure~\ref{fig:ha_vs_sed}
shows one of the main results of this work,
\textit{that between 2  and 10 Myr,
the clusters in M83 show a strong correlation between
their H$\alpha$ morphology and age}.
This suggests that the morphology of ionized gas alone
can be used as an age indicator. The best linear fit, shown in the figure,
is given by: 
$MC = (3.04\pm0.32) \times ~\mbox{log}(\tau/\mbox{yr}) -16.27\pm2.16$ (i.e., a 9 $\sigma$ correlation)
for $\log(\tau/\mbox{yr}) < 7$, where $MC$ is the morphological
category defined in Section~2. 
This relationship is only relevant when H$\alpha$ is
present (i.e., categories 3 and 4). 
A similar, flatter, and less well-defined relationship exists for category 5
and higher, as shown in the top figure:
$MC = (0.45\pm0.09) \times \log(\tau/\mbox{yr}) + 1.78\pm0.82$
(i.e., a 5 $\sigma$ correlation)
for $\log(\tau/\mbox{yr}) > 7$.

We note that a similar correlation exists between 
our morphological categories and ages estimated
by comparing color magnitude diagrams of individual stars with
stellar tracks for stars around some of our objects, as will
be discussed in a study of 50 regions in M83 by H.~Kim et~al.\ (2011, in preparation).

While the dispersion in the estimated SED ages within each morphological 
category is remarkably small (see Figure~\ref{fig:18plot} and Table 2),
a close inspection reveals four outliers in category~5b.
Three of these clusters (44034, 18032 and 74692) have
no associated H$\alpha$ emission, have relatively small pixel-to-pixel
flux variations (discussed in Section 4),
and have optical colors consistent with those predicted for
extinction-free intermediate age clusters, and therefore
almost certainly have ages
$\log(\tau/\mbox{yr})\approx8$. 
The SED fitting, however, erroneously assigned them lower
ages ($\log(\tau/\mbox{yr}) < 7$), and
higher extinctions (i.e., $E(B\!-\!V) = 0.28$, 0.36, 0.50 mag). 
The fourth cluster (10114) has significantly
bluer colors than the other clusters in category~5b.
In retrospect, this cluster probably belongs in morphological category 4a, 
since there does appear to be a small amount of  
H$\alpha$ emission associated with it.
In this case it is the morphological classification which appears to be
in error rather than the SED age estimate 
(the latter is $\log(\tau/\mbox{yr}) = 6.6$). 
Overall, we find good agreement between the previously determined
SED age and currently determined morphological category for
95\% of the clusters (i.e., 84 out of 88 cases), demonstrating that {\it both}
techniques are quite reliable.

\subsection{An Empirical Correlation Between H\boldmath{$\alpha$}
Bubble Size and Cluster Age}

While the good correlation between the morphological category and
SED age found in Figure~3 is encouraging, this approach has several limitations.
In particular, placing the clusters into the different categories
is subjective and hence not easily automated. 
An additional complication is that an isolated,
full 360 degree ring with a single dominant central cluster
is very rare. Partial rings, multiple  
loops, and multiple clusters in the region are more typical
situations.\footnote{We note that the term ``central'' cluster may be misleading in
some cases. There are many cases where the ring or shell is asymmetric, 
typically offset to the side where there is a more prominent dust lane.
Examples of this morphology for super shells in the Antennae galaxy and a discussion in terms of the ``blister'' model of Israel (1978) are provided in
Whitmore et~al.\ (2010).}

Here, we attempt to better quantify the relationship between 
H$\alpha$ morphology and cluster age by measuring
the radii of the most prominent ``coherent'' ring or partial shell 
of ionized gas associated with the cluster stars. 
While the measurement of H$\alpha$
bubble size  is also difficult to automate,
this more quantitative approach provides
the opportunity for improving the correlation still further, and for
exploring the impact that other physical parameters (e.g.,
cluster mass) have on this relationship.

Figure~\ref{fig:ha_vs_r} plots the
measured H$\alpha$ bubble size versus photometric
age for clusters younger than $\approx10$~Myr
(top panel), and
bubble size versus morphological category (bottom panel).
There is clearly a strong correlation which suggests that
H$\alpha$ morphology alone can be used as
an age indicator for clusters younger than $\approx10$~Myr.
This figure hints at other intriguing results as well.
The correlation between age
and bubble size is strongest for clusters younger than
$\log(\tau/\mbox{yr}) \approx 6.7$, with typical bubbles
growing from a few pc to $\approx20$~pc in size.
After this time, between ages $\log(\tau/\mbox{yr}) \approx 6.7$--6.9,
there is a large range observed in bubble size,
from $\approx20$~pc to nearly 80~pc, and no strong
dependence on age.
These trends are illustrated by the solid lines, which
represent simple linear fits to observations in the
two different regimes. The best linear fits for the top panel
are given by: 
$\log(\tau/\mbox{yr})  = (0.015\pm0.002) \times ~R (pc) + 6.46\pm0.03$
for R $<$ 20 pc, and 
$\log(\tau/\mbox{yr})  = (0.0014\pm0.0011) \times ~R (pc) + 4.37\pm0.22$
for R $>$ 20 pc. 
Similar fits are shown in the bottom panel, but numerical values are
not included here since the use of morphological category makes them more
qualitative. The large, open circles in the top panel of Figure~\ref{fig:ha_vs_r} 
show the clusters with $M < 1\times10^4~M_{\odot}$ and $\log(\tau/\mbox{yr}) > 6.7$. These  will be discussed further in Section 6.

\section{Surface Brightness Fluctuations as Age Indicators for \boldmath{$\tau \lea 100$}~Myr Clusters}

As discussed in Section~2, bright individual stars are expected to be
visible within
M83 clusters at ages younger than $\approx100$~Myr, giving the
clusters a mottled appearance with relatively large pixel-to-pixel flux
variations.
After this, the flux variations have mostly disappeared,
leading to a more uniform appearance among cluster stars.
Compare, for example, the images in categories 4b and 5a
in the top panel of Figure~\ref{fig:18plot},
where a number of bright individual cluster stars are clearly
visible, with the images
for categories 5b and 6, where individual cluster stars are no longer
observed. Figure 2 also shows a variety of cases where older clusters
appear as fuzzy objects while single 
individual stars are clearly observed within the young clusters.

Here, we use the strength of these surface brightness fluctuations
to develop a new method for estimating the ages of clusters in M83. 
This is reminiscent of the 
surface brightness fluctuation method used to estimate
distances to early-type galaxies (e.g., Tonry \& Schneider 1988),
but in our case the clusters are all at the same distance
and the surface brightness fluctuations are caused
by differences in age. One must exercise caution when using this
technique for other galaxies, since the relationships
derived below for M83 would need to be renormalized for galaxies
at different distances.

We first need to remove the overall radial gradient in the cluster
luminosity profile, since this gradient typically dominates
over the peak-to-peak variations in pixel flux within the cluster.
We accomplish this by creating a 
median divided image (with a $3\times3$ smoothing box size),
which effectively flattens the radial profile of
the cluster, leaving behind primarily the smaller
scale pixel-to-pixel variations that we are
interested in here. We then measure
the RMS scatter in the fluxes of pixels located
within a 10 pixel radius. A more sophisticated
method is being developed by  
C.~Kaleida et~al.\ (2011, in preparation) where the RMS variations are measured within
the half-light or effective radius R$_\mathrm{eff}$ of the cluster, hence
accounting for clusters of different sizes in a more systematic manner.

Figure~\ref{fig:sbf} shows a comparison of the F555W image and the
median divided F555W image in a region including clusters younger
than $\log(\tau/\mbox{yr}) \approx 7.0$ (blue circles and labels) 
and older than this (red circles and labels). 
The RMS measurements are indicated in the bottom panel. 
One can easily see that old clusters (e.g., nos.~65 and 70)
nearly disappear from the median divided image, resulting in low
values of the RMS, while bright, individual stars are observed 
in the younger clusters (e.g., nos.~32 and 49), resulting in a higher RMS. 
Small, barely visible dust lanes also contribute to larger values of
RMS for the young clusters.
In a few very concentrated clusters there are
small artifacts at their centers, for example in cluster no. 64.
However, the RMS is dominated by
pixels outside of this central region, hence this is a relatively
small effect. We note that for four clusters, identified in Table~1,
we have used a slightly smaller
radius to avoid single bright stars that are almost certainly unrelated to the
cluster (e.g., nos.~66, 69, 70 and 79). 

In the upper panel of Figure~\ref{fig:m83_rms_2plot}
we plot the RMS in the pixel-to-pixel flux measured in the $V$ band
versus the morphological category.
The resolution is better here for clusters younger than $\approx 10^7$~yr (i.e., category 3 and 4 objects).
In the lower panel of Figure~\ref{fig:m83_rms_2plot} we plot the RMS vs.\ the SED age, which gives better resolution for clusters older than $\approx10^7$~yr.
These figures show the second main result of our work, \textit{that 
older clusters have smaller RMS values of the pixel-to-pixel flux variations.} 
However, we also note, especially in the top panel,
that the RMS is degenerate, and increases in strength over the age range
2--7 Myr (i.e., morphological categories 3 and 4), peaks in category 4,
corresponding to an age of approximately 5--10~Myr,
and decreases in strength for clusters with older ages
and in later categories.

There are several possible explanations for this behavior.
The first, and probably most important effect,  
is that the brightest stars (i.e., red and blue super giants) do not appear until
about  5 Myr, and are largely gone by about 20 Myr.
Two other evolutionary effects may potentially contribute to the
low RMS measured for the very young clusters. 
(1)~The central part of the cluster, where it is difficult to detect individual
stars due to the high background, might
emerge from its dust cocoon before the outskirts. (2)~Clusters may start out very
compact and expand with age. Hence crowding may make it impossible
to detect individual stars at very early ages. 
There is both theoretical and observational support for 
both of these conjectures, as discussed below.

These trends are illustrated by the lines in Figure~\ref{fig:m83_rms_2plot}, 
which represent linear fits to observations. The best fits for the top panel
are given by: 
$MC = (8.57\pm1.61) \times~RMS + 2.85\pm0.18$ (i.e., a 5 $\sigma$ correlation) for category $<$4.7,
and $MC = (-7.23\pm1.25) \times~RMS + 5.79\pm0.11$ (i.e., a 6 $\sigma$ correlation) for category $\geq$4.7.

For the bottom panel, only the trend for older clusters is shown, with
$\log(\tau/\mbox{yr}) = (20.43\pm5.43) \times~RMS + 9.59\pm0.31$ 
for $\log(\tau/\mbox{yr}) > 7$.
We will better quantify the correlation between
cluster age and the RMS in pixel-to-pixel fluxes in a future
paper using a significantly larger number of clusters.
We note that these relationships will be different for
galaxies at different distances, and in different filters.
For example, fluctuations will generally be larger for older clusters
(100--1000~Myr) in the near-IR due to the presence of asymptotic 
giant branch (AGB) stars.

Values of RMS pixel-to-pixel flux variations for each cluster
are included in Table 1, and the mean values for 
clusters in each morphological category
are included in Table 2. The RMS method is particularly
promising for estimating the ages of clusters between
10 and 100 Myr, a range which can be
somewhat difficult for the SED method because the integrated colors
of clusters loop back on themselves during this period. 
Clusters in this age range can be identified initially by their lack of 
ionized gas.

It is instructive to look at the outliers in 
Figure~\ref{fig:m83_rms_2plot}, just as we did for 
the correlation between SED age and morphological category in Section~3.1. 
In the bottom panel the cluster
with RMS~= 0.03 and $\log(\tau/\mbox{yr}) = 6.8$ (ID~= 44034) 
is an apparent outlier. It is also one of the sources which we believe,
based on the discussion in Section~3.1,
has an incorrect SED age estimate. Based on its measured colors and 
assuming $E(B\!-\!V) = 0.0$ instead of 0.28, this cluster has a likely
age of $\log(\tau/\mbox{yr}) \approx 7.8$, more appropriate for it's low 
RMS  value. Similarly, many of the data points just to the right of this one 
are the same category 5b outliers discussed in Section~3.1 that have 
suspect ages. Other data points in the same region
of the $\log(\tau/\mbox{yr})$ versus RMS
plot are located in the large bubble just below
the nucleus of M83, where it is difficult to measure the RMS 
accurately due to the very high background. 
We note that the four clusters with RMS~$< 0.07$ 
and $\log(\tau/\mbox{yr}) < 6.5$  are all in morphological  
category 3, and are largely responsible for the decline in 
the measured RMS for the youngest clusters.

Many of the trends discussed above are seen in Table 2,
which lists the mean values of several parameters 
as a function of morphological category. In particular, the 
trends in $E(B\!-\!V)$, $\log(\tau/\mbox{yr})$, and H$\alpha$ shell radius (when present) are clearly evident. In addition, the double-valued nature 
of pixel-to-pixel RMS values is evident, with a peak at category 4a. 
One correlation that was mentioned briefly is the trend of increasing cluster 
size with age, quantified by the increase in the 
concentration index (CI) from values around 2.5 for Categories 3 and 4a to $>3.0$
for Categories 5b and 6. We find the same trend in our M51 data, which
is briefly discussed in the Appendix. 
This effect appears to be real, and is likely related
to the rapid expansion of the clusters (see, for example, Mackey \& Gilmore 2003; Bastian et~al.\ 2009; Pfalzner 2009).
Several different physical mechanisms may be responsible for this expansion, 
including the expulsion of leftover ISM due to feedback from massive
stars (e.g., Goodwin \& Bastian 2006; Baumgardt \& Kroupa 2007), and heating by binary stars and stellar mass black holes (e.g., Mackey et~al.\ 2008; see Portegies Zwart et~al.\ 2010 for a review of this subject).
The early expansion of star clusters will be discussed in more detail in 
a future paper (R.~Chandar et~al.\ 2011, in preparation).

\section{``Single Star'' HII Regions}

While the dominant sources of ionizing flux responsible for the 
H$\alpha$ emission in M83 are
massive, young star clusters, which were studied in the previous
sections, there is also a population of compact HII regions 
which are ionized by what appear to be single stars.
In this section we identify and study
a sample of 22 HII regions with very small H$\alpha$ radii and 
an unresolved central point source (based on their CI,
i.e., the magnitude difference between 0.5 and
3 pixel radii; see Chandar et~al.\ 2010). Color images of the selected 
sources are shown in Figure~\ref{fig:target_mosaic}. Only the
brighter candidates have been retained for this first exploratory study; the 
sample could be increased by a factor
of two or more if fainter, less distinct objects were included. 
The measured colors of these sources, shown
in Figure~\ref{fig:mini_cmd}, either coincide with the 
colors predicted for the bluest (youngest)
stars in the upper left portion of the diagram, 
or are found downstream along the reddening vector.
Basic properties of these sources,
including positions, $M_V$, CI, colors and measured
radii of the H$\alpha$ bubble, are compiled in Table~3.

We note that it is unlikely that all of these objects are actually
individual massive stars, hence the use of quotes around ``Single Star.'' 
Some, or even most of these may be a dominant star among
a close grouping of stars which
are either too close to the primary source, or too faint, to
be detected. 
What we can say is that a single (or very close binary) star
dominates the light profile, resulting in a CI
that is indistinguishable from a single star.

If we assume that all of these objects have a similar
age, and that the distribution in the two-color
diagram is primarily due to reddening, we
can correct for the effects of reddening and extinction.
We show the corrected photometry in a color magnitude diagram
in the left panel of Figure~\ref{fig:mini_cmd}.
Here, we have assumed an intrinsic color of $U\!-\!I = -2.2$,
the color of the bluest object, and solve for the $V\!-\!I$
color excess of each HII region.
We find that this procedure moves most of
the objects to a very young isochrone (4 Myr
is shown), although three sources scatter to the left
of the models. These may be especially young stars, or the offset
may  result from observational uncertainties, since these are the three
most reddened sources, and hence have the largest corrections.

While many of these single-star HII (i.e., SSHII) regions  are 
found near large regions of recent star formation,
several of them are quite isolated (see Figure 3 from Whitmore 2010), 
raising the possibility that these massive stars formed
in the field rather than in clusters or associations.
For example, five of  the 22 SSHII regions are $\approx$1 kpc
away from any region of active star formation. 
If these are stars that have been dynamically ejected
from their birthsites, i.e., runaways, they must have velocities
$\approx$200 km/s. However, they do not appear to have the prominent bow-front morphologies typical of many runaway stars in
the Galaxy and Magellanic Clouds (e.g., Gvaramadze et~al.\ 2011).
Velocity measurements are needed 
to establish if these massive stars formed in the field or 
are runaways from larger star-forming regions.

\section{Discussion}

As summarized in the Introduction,
the spatial relationship between young $\tau \lea 10^7$~yr
star clusters and their HII regions has been previously studied
both observationally and theoretically. To our knowledge however, no study
has yet systematically measured the sizes of H$\alpha$ bubbles in a 
nearby galaxy and correlated these sizes with the properties of their 
ionizing star clusters, as we have done here.
We have demonstrated that in M83 there is a good correlation between
bubble size and age for clusters younger than 
$\log(\tau/\mbox{yr}) \approx 6.7$, and that clusters with ages 
$\log(\tau/\mbox{yr}) \approx 6.7$--6.9 have a larger range in bubble size.
This suggests that some of the expanding bubbles at these older
ages may have effectively ``stalled.''

Theoretical work suggests that bubble size should depend not
only on the ambient pressure in the ISM and
the age of the cluster, but also on cluster mass.
To look for evidence of a mass-dependence on bubble size,
we compare the sizes measured for H$\alpha$ bubbles 
associated with 
$\log(\tau/\mbox{yr}) = 6.7$--6.9 clusters that are
more and less massive than $10^4~M_{\odot}$.
At these ages the bubbles are presumably approaching their stall radius,
and the effect of cluster mass should be more apparent.
The large, open circles in the top panel of Figure~\ref{fig:ha_vs_r} 
show the sizes of clusters with
$M < 1\times10^4~M_{\odot}$ and $\log(\tau/\mbox{yr}) > 6.7$.
While clusters at high and low masses have an overlapping range in cluster
size, there is a tendency for more massive clusters to
have larger bubbles, with median sizes of 18~pc and 41~pc for 
$M < 10^4~M_{\odot}$ and $M \geq 10^4~M_{\odot}$, respectively.
A formal fit of H$\alpha$ radius versus $\log(M/M_{\odot})$
for these clusters gives a slope of $0.006\pm0.005$, suggesting 
that there may be a weak correlation between bubble size and mass.
Future studies that include a larger number of sources, selected
in a more systematic way, are needed to confirm this result.

However, other effects may also be important. For example,
stochasticity in the number of massive stars formed in lower mass
clusters can strongly affect their integrtated colors and hence their
estimated ages. In this case, a lower mass cluster can have an
intrinsically redder color than a higher mass cluster of the same age
(e.g., Fouesneau \& Lancon 2010), and hence an older estimate for the
age of the cluster.

In addition to the correlation between H$\alpha$ bubble size and
cluster age (and possibly mass), we also found that the RMS variations 
in pixel-to-pixel brightness correlates with cluster age.
This result has applications beyond star clusters, and can 
be used to constrain the ages of stellar populations in general.
For example, portions of M83 itself that contain young stars and star
clusters, such as along the spiral arms, have larger 
pixel-to-pixel flux variations than portions of M83 dominated
by older stars, such as between the arms
(see H.~Kim et~al.\ 2011, in preparation for a discussion).
This is also true in large portions of M82 and in portions of 
the tidal tails of galaxy mergers, where regions
dominated by intermediate age star clusters ($\approx$100~Myr)
have small fluctuations in the surface brightness of the field stars. 

\section{Summary and Conclusions}

We have used observations 
taken with the newly installed WFC3 camera
on-board the {\em Hubble Space Telescope} to develop
two independent methods for age-dating young
star clusters in the nearby spiral galaxy M83.
Our primary results are summarized below.

1. A working classification system, largely based  on 
H$\alpha$ morphology and pixel-to-pixel flux variations, 
was developed to map an observed age sequence onto a 
proposed sequence of cluster evolution.
The underlying evolutionary picture includes the formation of dense cores in
giant molecular clouds (GMCs); a stage where the young stars are completely
obscured by their dust cocoon; an emerging stage where the clusters
become visible in the IR, and then in the optical as stellar winds and
supernovae blow away the dust; a stage where an expanding H$\alpha$
bubble is blown and the existence of very bright young stars leads to
large pixel-to-pixel flux variations; and later stages with no 
evidence of H$\alpha$ and diminishing pixel-to-pixel flux variations. 

2. We found that H$\alpha$ morphology, i.e., the size of the ionized gas
bubble, provides a viable method for age-dating
clusters in the range 1--10~Myr.
This method is based on the simple premise that the gas in very young
($\tau <$~few Myr) clusters is largely coincident with the cluster
stars, is in a small, ring-like structure surrounding the stars in
slightly older clusters since the winds from massive stars have had
time to push out the natal gas (e.g., $\tau \approx 5$~Myr), and is in
a larger ring-like bubble for still older clusters (i.e., $\approx$5--10~Myr). 
If no H$\alpha$ is associated with a cluster it is generally older
than $\approx$10~Myr.

We first made  {\it qualitative\/} estimates based on the classification 
scheme outlined above, and find that the ages of the clusters,
as determined from the SED method described in Chandar et~al.\ (2010),
correlate well with the morphological categories, with a 
scatter of $\approx$0.1 in $\log(\tau/\mbox{yr})$
within each category for the clusters with H$\alpha$ emission, and
a scatter of $\approx$0.5 in $\log(\tau/\mbox{yr})$ for the older clusters.
We then {\it quantified\/} this technique by correlating the measured 
radii of the most conspicuous H$\alpha$-emitting ring or shell which
appears to be physically related to the cluster with the SED ages 
determined in Chandar et~al.\ (2010). We found tentative evidence 
for a weak correlation between bubble size and cluster mass, but a 
larger, more objectively selected sample will be required to confirm this.

3. We then used pixel-to-pixel flux variations to age-date
clusters.  This technique is based on the fact that individual stars
are bright enough to be visible within clusters when they are young
(e.g., M$_V < -3.5$, the approximate detection level, for the
brightest stars with ages $<$100 Myr), leading to relatively large
pixel-to-pixel variations in flux.  The strength of the fluctuations
peaks in clusters with ages of $\approx$5--10~Myr, presumably because
this is when the brightest stars (e.g., red and blue super giants) appear.  
The number of luminous, evolved stars falls off for both younger and 
older clusters.  This degeneracy in age can be broken using other age 
indicators such as the H$\alpha$ morphology.  The technique is especially 
useful for identifying clusters older than 100 Myr.

4. A by-product of this study was the identification of 22
``single-star'' HII regions in M83.  By assuming that all of these
objects have a similar age, and that the distribution in the two-color
diagram is primarily due to reddening, we corrected for the effects of
reddening and extinction.  We found that this procedure moves most of
the objects to a very young isochrone with an age approximately 4~Myr. 
Some of these massive stars are located far from any star-forming region,
indicating that they either formed in the field or were dynamically ejected 
from their birthsites at very high velocities. These SSHII regioins will be 
studied in more detail in H.~Kim et~al.\ (2011, in preparation).

In the future, we will extend the classification system into the near-IR (i.e., categories~1 and 2) using our J and H observations. We will also 
calculate the energy budget of cluster stars and compare  with  physical
properties of the clusters and the ISM. Finally,  we will extend this analysis 
to other galaxies in the ERS1 sample (including ``low pressure'' systems 
such as the dwarf starburst galaxy NGC~4214) in order to determine whether 
the relationships are universal or are strongly dependent on environment.

\bigskip 

\acknowledgments

\begin{figure}
\includegraphics[angle=-90,width=\textwidth]{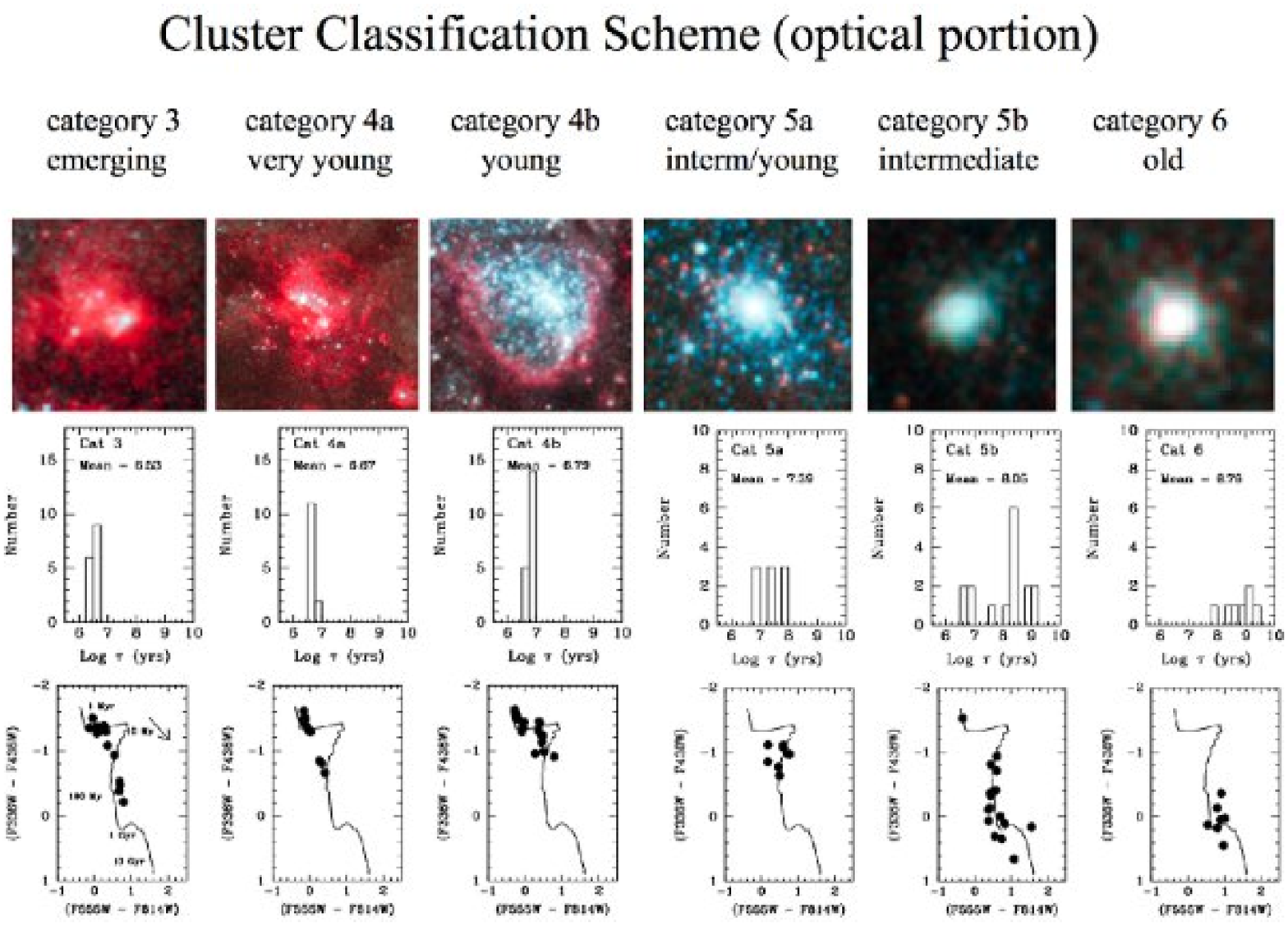}
\caption{Examples of sources in the various
categories defined in Section~2 are shown along the top row; 
histograms of the SED age estimates for the clusters in Table 1 (from Chandar
et~al.\ 2010) are in the middle row, 
and two-color diagrams comparing the (measured) integrated colors
of the clusters with predictions for a twice solar metallicity BC09 model
(G.~Bruzual \& S.~Charlot 2009, private communication;  see also Bruzual \& Charlot 2003) are shown along the bottom row.
}
\label{fig:18plot}
\end{figure}

\begin{figure}
\epsscale{1.0}
\plotone{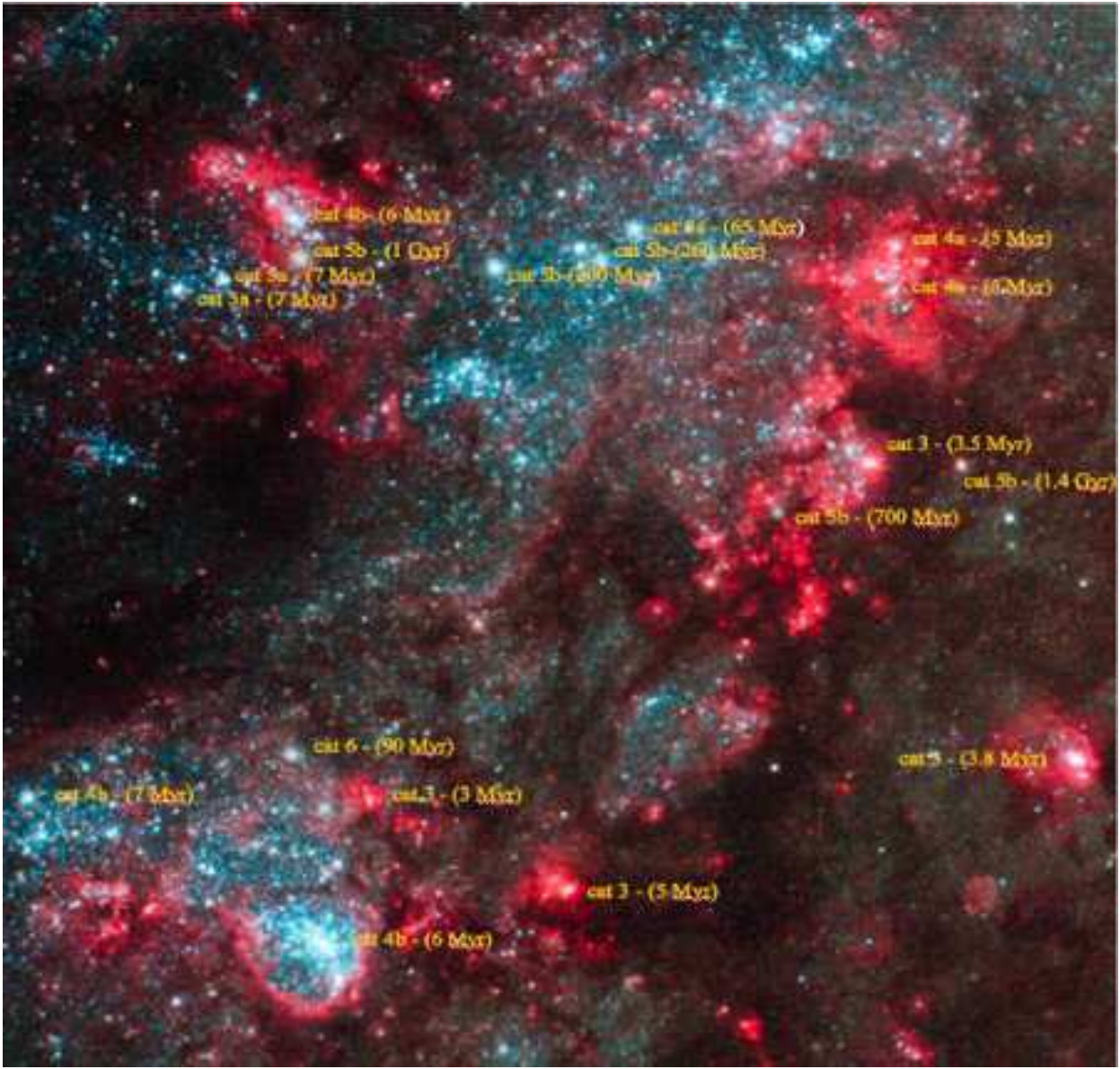}
\caption{Portion of the M83 field  ($1.0\times1.0$ kpc) showing the good correlation between morphological categories determined in this paper and 
SED age estimates from Chandar et~al.\ (2010). 
}
\label{fig:large_image}
\end{figure}

\begin{figure}
\epsscale{0.5}
\plotone{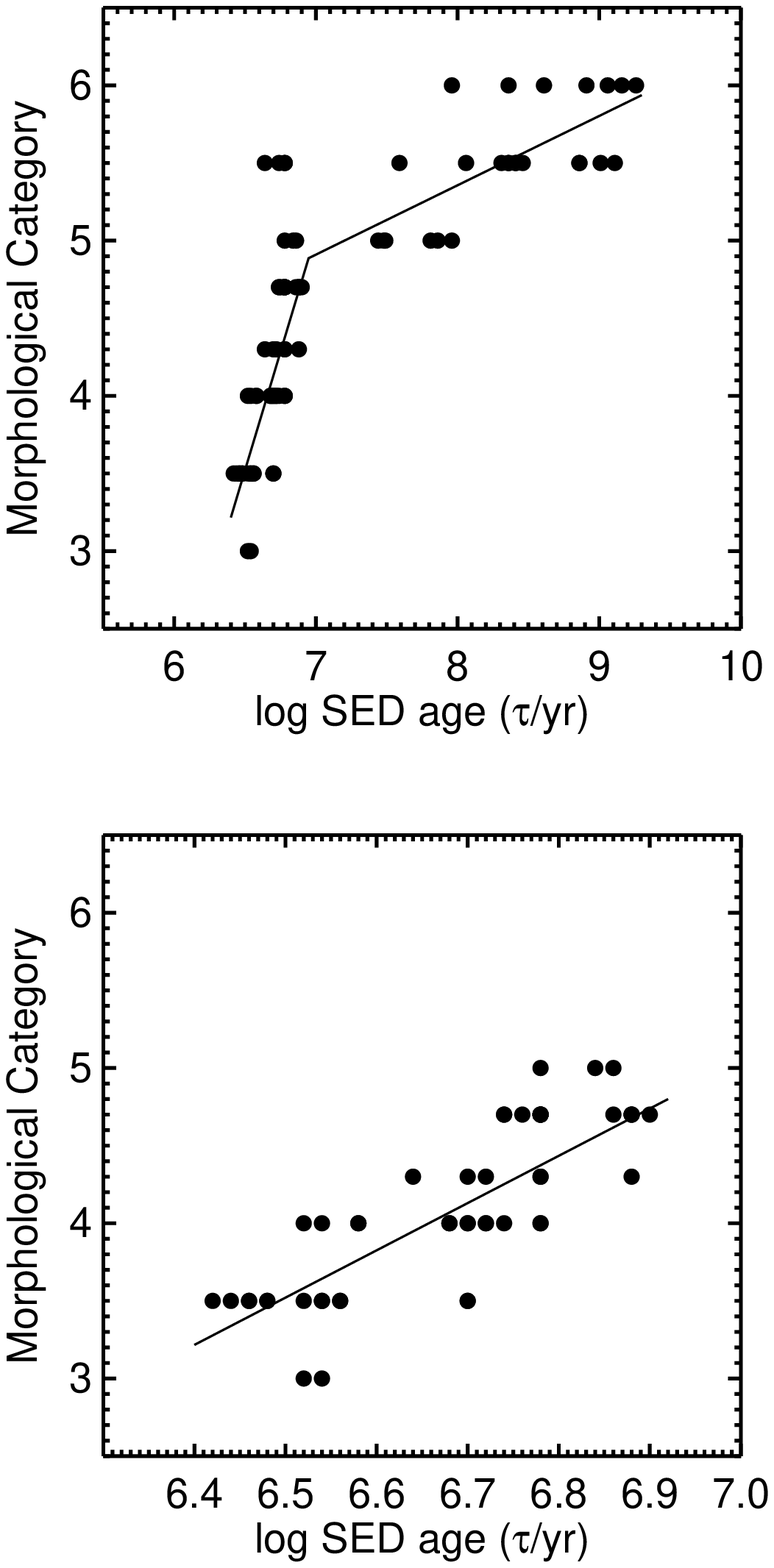}
\caption{Plot of morphological categories versus log SED age estimates. Note how the scatter increases with age and that the relationship is nonlinear, as 
expected since H$\alpha$ morphology evolves quickly between the ages of a
few to 10 Myr. The bottom panel shows an enlargement for the younger
ages. The best linear fit is shown in both panels. 
}
\label{fig:ha_vs_sed}
\end{figure}

\begin{figure}
\epsscale{0.5}
\plotone{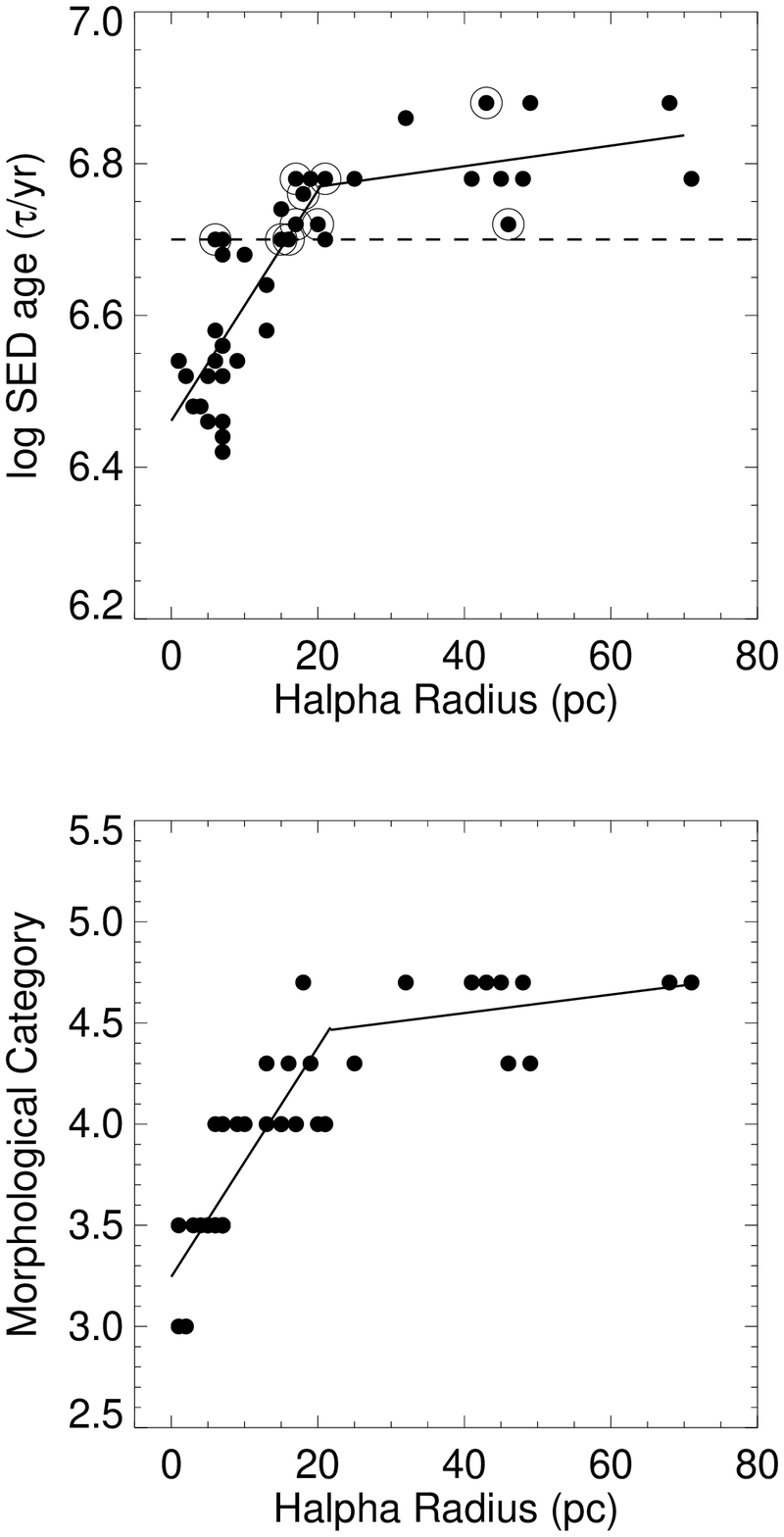}
\caption{
Plot of the H$\alpha$ bubble radius vs log SED ages in the upper panel
and morphological category in the lower panel. The correlation is similar
to the relationship shown in Figure~3, but with fewer outliers and hence 
smaller scatter. The large, open circles in the top panel show the sizes of clusters with $M < 1\times10^4~M_{\odot}$ and 
$\log(\tau/\mbox{yr}) \geq 6.7$, as discussed further in Section~6. 
}
\label{fig:ha_vs_r}
\end{figure}

\begin{figure}
\includegraphics[angle=-90,width=\textwidth]{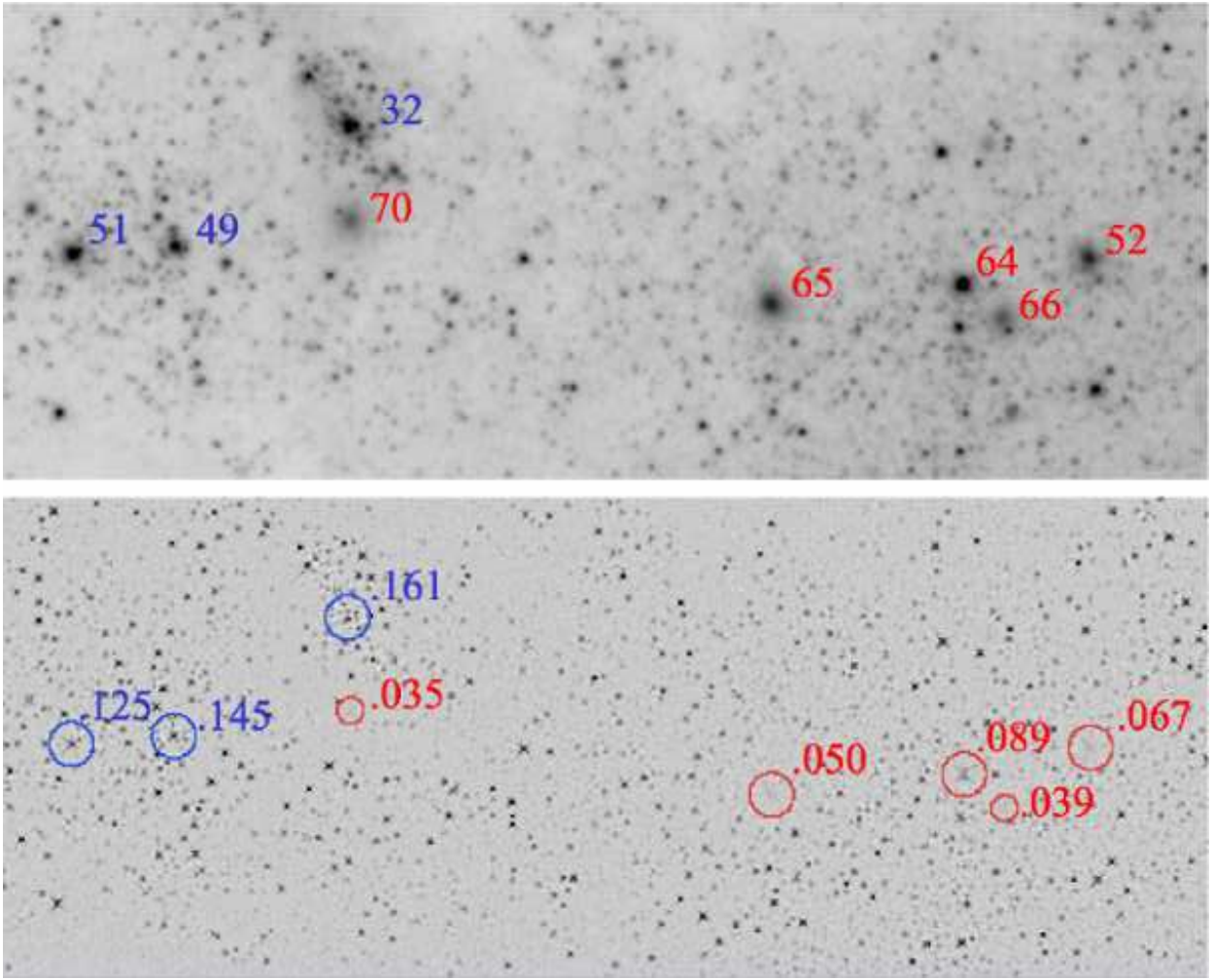}
\caption{Top: F555W image (log stretch) for a portion of M83 (i.e., the upper left of Figure~2). Blue labels are used for clusters younger than 10 Myr while red labels are used for clusters older than 10 Myr. Bottom: median divided image for the same region with the aperture size and rms of the pixel-to-pixel flux variations shown.
}
\label{fig:sbf}
\end{figure}

\begin{figure}
\epsscale{0.50}
\plotone{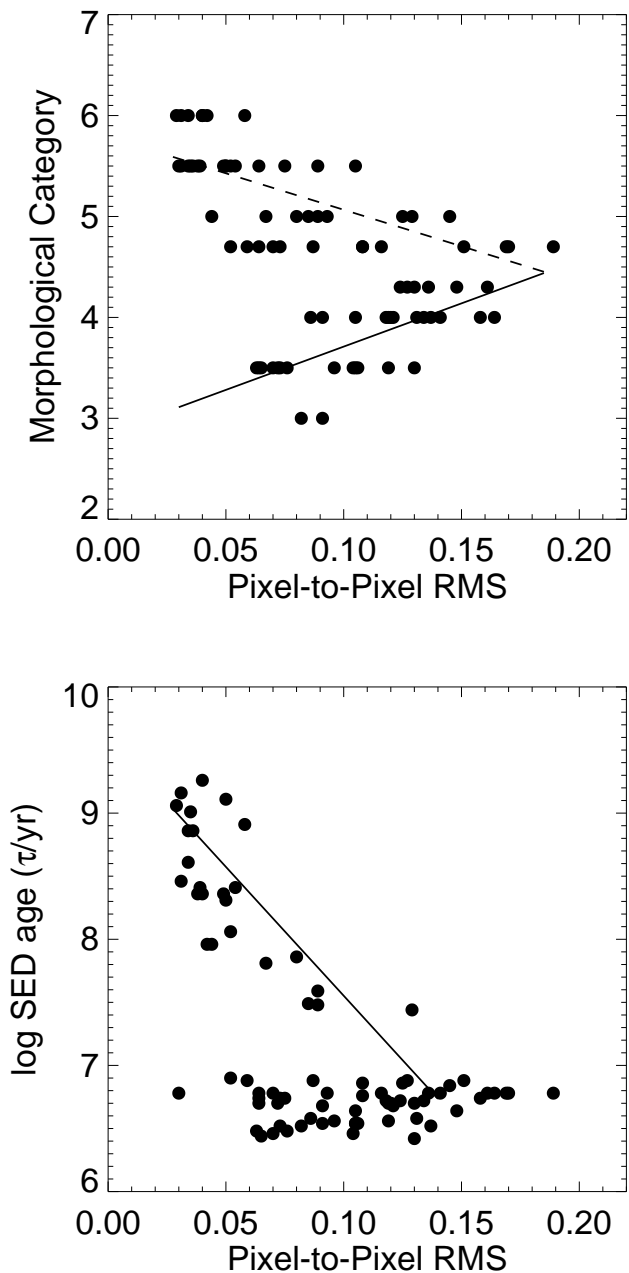}
\caption{
Plot of the pixel-to-pixel flux variation (RMS) versus log SED age in the lower panel, and morphological category in the upper panel. The highest values of the
RMS are seen in the 5--10 Myr range (category 4b), with lower values for both younger and older clusters. Note that nearly all the clusters with RMS $<$ 0.05 have ages greater than 100 Myr. 
}
\label{fig:m83_rms_2plot}
\end{figure}

\begin{figure}
\epsscale{0.80}
\plotone{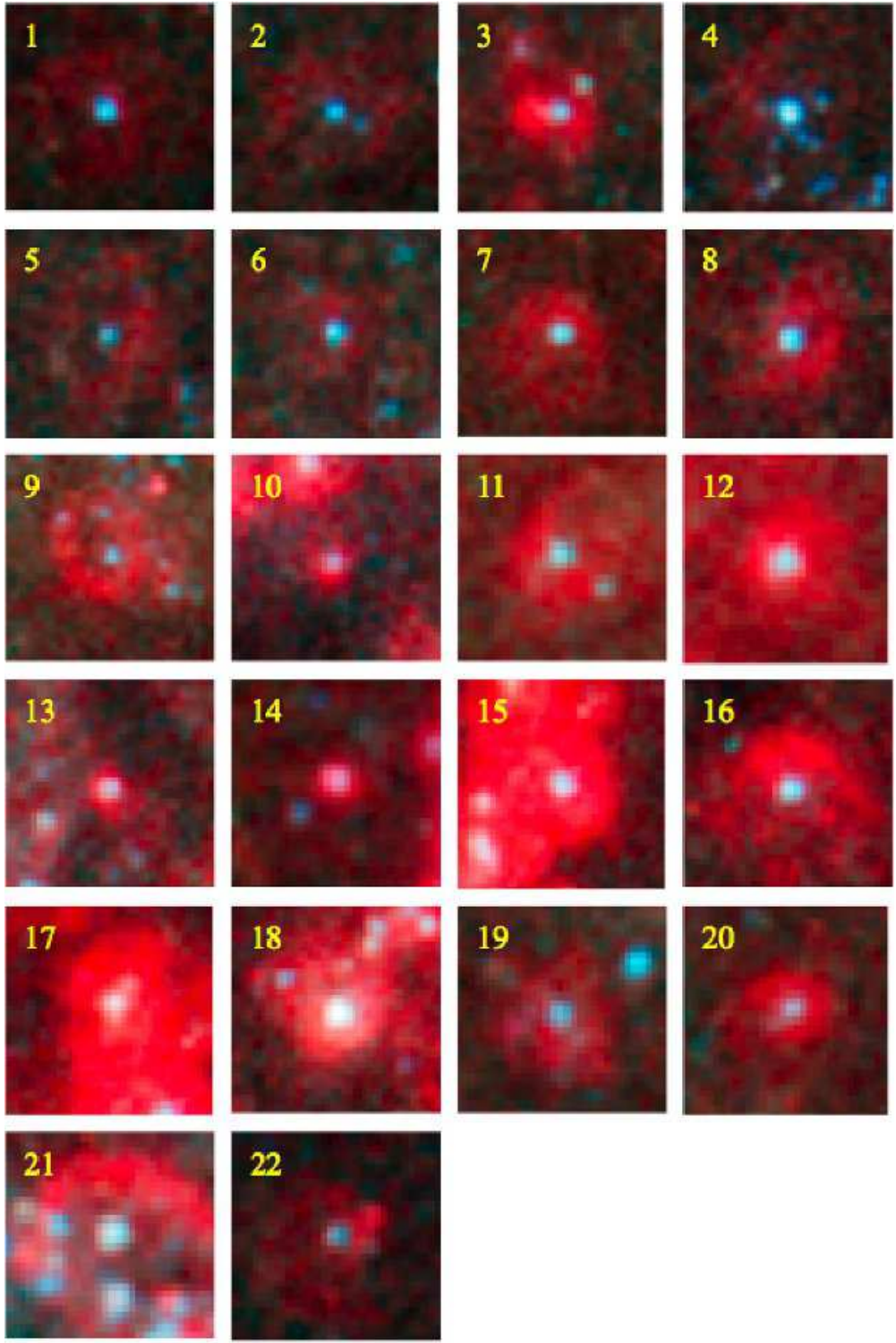}
\caption{Mosaic of the 22 SSHII region candidates. 
}
\label{fig:target_mosaic}
\end{figure}

\begin{figure}
\epsscale{0.90}
\plotone{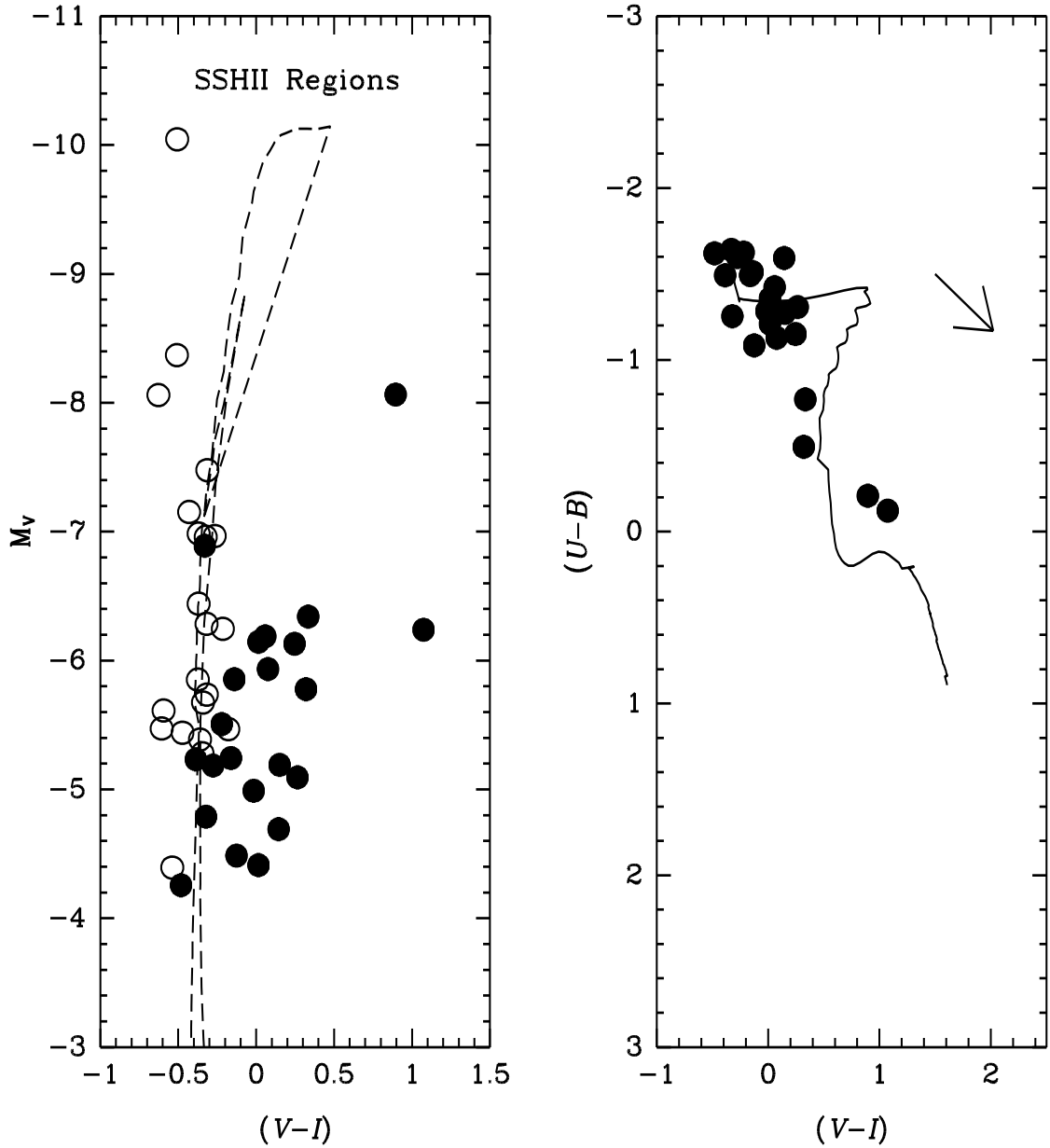}
\caption{Left: Color-magnitude diagram for the SSHII region sample.
The filled circles show the observed values while the open circles show
the corrected values based on the extinction derived from the
color-color diagram (see the text). 
Right: Color-color diagram for the SSHII region sample.
}
\label{fig:mini_cmd}
\end{figure}

We thank Zolt Levay for making the color images used in 
Figures~\ref{fig:18plot} and \ref{fig:large_image}.
This paper is based on observations taken with the NASA/ESA
{\em Hubble Space Telescope} obtained at the Space Telescope
Science Institute, which is operated by AURA, Inc., under 
NASA contract NAS5-26555. 
The paper makes use of Early Release Science observations made by the 
WFC3 Science Oversight Committee.
We are grateful to the Director of STScI for
awarding Director's Discretionary time for this program.
R.~C.\ is grateful for support from the NSF through CAREER award 0847467.
This research has made use of the NASA/IPAC Extragalactic Database 
(NED), which is operated by the Jet Propulsion Laboratory, California 
Institute of Technology, under contract with NASA.

{\it Facilities:} \facility{HST}.

\centerline{\bf Appendix}

\begin{figure}
\epsscale{1.0}
\plotone{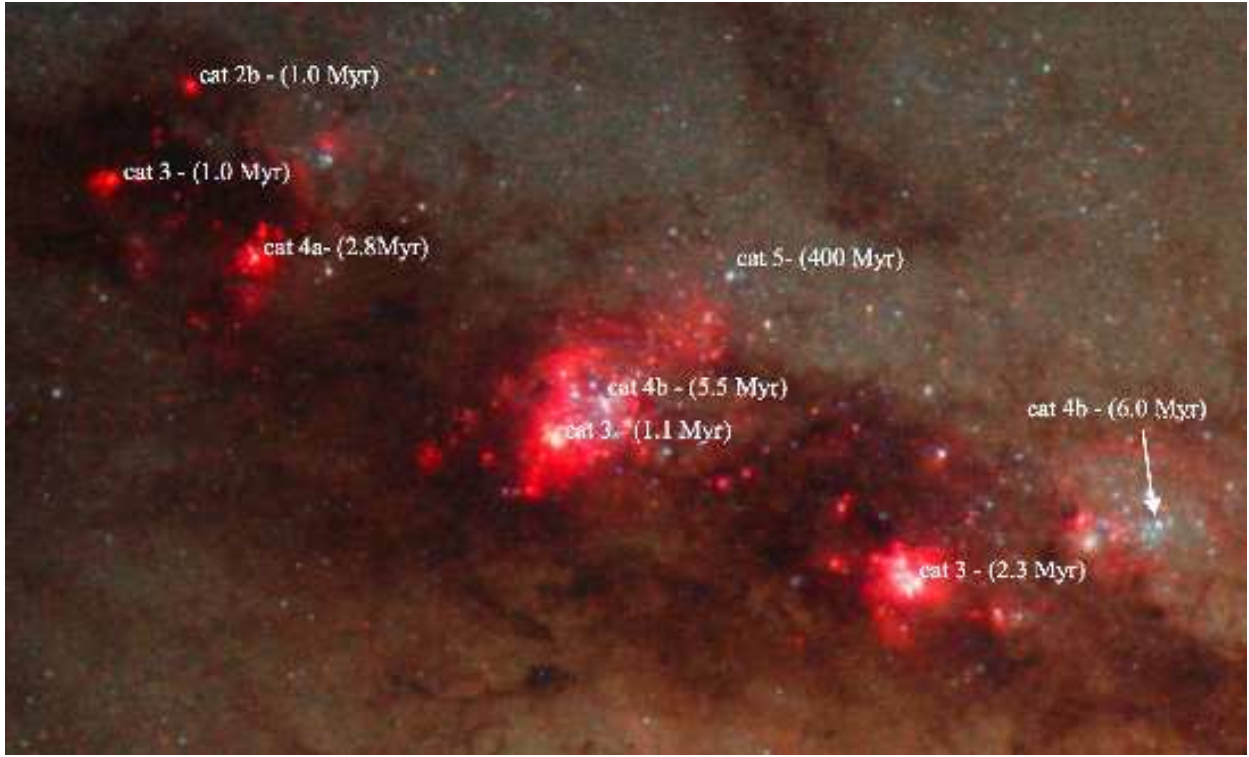}
\caption{Portion of the M51 image showing the good correlation between
morphological categories and SED age estimates (R.~Chandar et~al.\ 2011, in preparation). Note the similarity with Figure~\ref{fig:large_image}, implying  that this method of age dating will work to at least twice the distance of M83. 
}
\label{fig:m51_ha_morph_1}
\end{figure}

\begin{figure}
\plotone{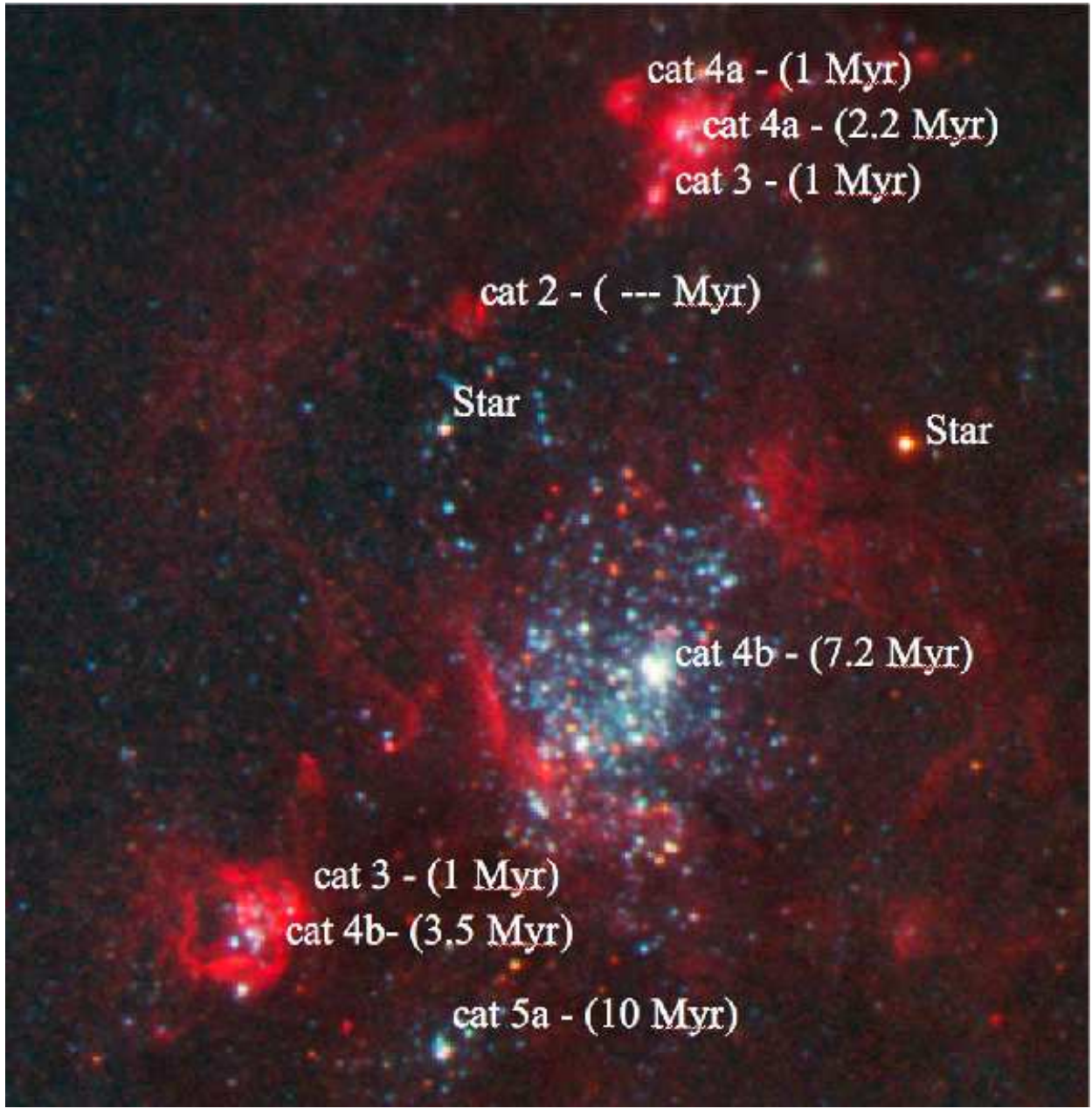}
\caption{Similar to Figure~\ref{fig:m51_ha_morph_1}, but for a different region.
}
\label{fig:m51_ha_morph_2}
\end{figure}

The spiral galaxy M51 was originally used to perform a pilot study
to test how well H$\alpha$ morphology can be used to estimate
cluster ages. Examples of the results are included in 
Figures \ref{fig:m51_ha_morph_1} and \ref{fig:m51_ha_morph_2}. 
These are similar to Figure~\ref{fig:large_image}  for M83 and show that even
at twice the distance of M83 the quality of the images are comparable, and the 
morphological categories match SED age estimates at a similar level. 
A more complete analysis will be included in a future paper (R.~Chandar 
et~al.\ 2011; in preparation).

\clearpage

\clearpage

\begin{deluxetable}{rrllrcrrccclcc}
\tabletypesize{\scriptsize}
\rotate
\newcommand\cola {\null}
\newcommand\colb {&}
\newcommand\colc {&}
\newcommand\cold {&}
\newcommand\cole {&}
\newcommand\colf {&}
\newcommand\colg {&}
\newcommand\colh {&}
\newcommand\coli {&}
\newcommand\colj {&}
\newcommand\colk {&}
\newcommand\coll {&}
\newcommand\colm {&}
\newcommand\coln {&}
\newcommand\eol{\\}
\newcommand\extline{&&&&&&&&&&&&&\eol}

\tablecaption{Parameters for 88 Selected Clusters\label{tab:table_1}}
\tablewidth{0pt}
\tablehead{
\colhead{\#} & \colhead{ID} & \colhead{RA} & \colhead{DEC} & \colhead{M$_V$\rlap{$^a$}} & \colhead{CI\rlap{$^b$}} & \colhead{$U\!-\!B$\rlap{$^a$}} & \colhead{$V\!-\!I$\rlap{$^a$}} & \colhead{$E(B\!-\!V$)} & \colhead{Log Age} & \colhead{Mass} & \colhead{Cat\rlap{$^c$}} & \colhead{R(H$\alpha$)} & \colhead{RMS\rlap{$^d$}} \\
 & & & & (mag) & (mag) & (mag) & (mag) & (mag) & (yr) & (M$\odot$)  & & (pc)   }

\startdata

\cola     1\colb   65815\colc   204.244637\cold $-$29.8507043\cole    $-$6.10
\colf     2.57\colg    $-$1.35\colh    $-$0.12\coli    0.24\colj    6.52\colk      1.1E3
\coll      3\colm        \phn2\coln    0.082\eol
\cola     2\colb   63611\colc  204.2426354\cold $-$29.8515924\cole    $-$6.42
\colf     2.35\colg    $-$1.27\colh     0.07\coli    0.34\colj    6.54\colk      2.1E3
\coll      3\colm        \phn1\coln    0.091\eol
\cola     3\colb   41736\colc   204.251687\cold $-$29.8640556\cole    $-$8.70
\colf     2.68\colg    $-$0.21\colh     0.80\coli     0.50\colj     6.7\phn\colk      2.6E4
\coll      3\colm        \phn7\coln    0.064\eol
\cola     4\colb   39744\colc  204.2524132\cold $-$29.8651911\cole   $-$11.03
\colf     2.58\colg    $-$1.33\colh     0.23\coli     0.30\colj    6.56\colk      1.2E5
\coll      3\colm        \phn7\coln    0.119\eol
\cola     5\colb   49651\colc  204.2564416\cold $-$29.8586912\cole    $-$7.17
\colf     2.50\colg    $-$0.38\colh     0.68\coli     0.50\colj    6.48\colk      7.1E3
\coll      3\colm        \phn3\coln    0.063\eol
\cola     6\colb   32461\colc  204.2880034\cold $-$29.8692326\cole    $-$7.67
\colf     2.62\colg    $-$1.36\colh     0.30\coli    0.44\colj    6.52\colk      8.6E3
\coll      3\colm        \phn5\coln    0.073\eol
\cola     7\colb   53072\colc  204.2889299\cold $-$29.8569125\cole    $-$6.64
\colf     2.61\colg    $-$0.93\colh     0.56\coli     0.50\colj    6.48\colk      4.2E3
\coll      3\colm        \phn4\coln    0.076\eol
\cola     8\colb   93801\colc  204.2795357\cold $-$29.8392001\cole    $-$7.10
\colf     2.50\colg    $-$1.37\colh     0.08\coli    0.32\colj    6.54\colk      3.6E3
\coll      3\colm        \phn1\coln    0.106\eol
\cola     9\colb   50287\colc  204.2865616\cold $-$29.8582915\cole    $-$7.35
\colf     2.66\colg    $-$0.54\colh     0.69\coli     0.50\colj     6.70\colk      7.4E3
\coll      3\colm        \phn6\coln    0.072\eol
\cola    10\colb   59214\colc  204.2818977\cold $-$29.8537395\cole    $-$7.56
\colf     2.53\colg    $-$1.08\colh     0.36\coli     0.50\colj    6.54\colk      9.3E3
\coll      3\colm        \phn6\coln    0.105\eol
\cola    11\colb   24767\colc  204.2493318\cold $-$29.8729735\cole    $-$6.89
\colf     2.32\colg    $-$0.46\colh     0.70\coli     0.50\colj    6.56\colk      5.1E3
\coll      3\colm        \phn7\coln    0.096\eol
\cola    12\colb   58013\colc  204.2475174\cold $-$29.8543366\cole    $-$6.49
\colf     2.43\colg    $-$1.29\colh     0.15\coli    0.42\colj    6.46\colk      5.4E3
\coll      3\colm        \phn7\coln    0.104\eol
\cola    13\colb   52726\colc  204.2458956\cold $-$29.8570619\cole    $-$8.13
\colf     2.78\colg    $-$1.50\colh    $-$0.06\coli    0.26\colj    6.42\colk      1.2E4
\coll      3\colm        \phn7\coln    0.130\eol
\cola    14\colb   12721\colc  204.2908632\cold $-$29.8785688\cole    $-$5.98
\colf     2.43\colg    $-$1.30\colh     0.32\coli     0.50\colj    6.46\colk      4.3E3
\coll      3\colm        \phn5\coln    0.070\eol
\cola    15\colb   19354\colc  204.2525635\cold $-$29.8755537\cole    $-$6.06
\colf     2.83\colg    $-$1.39\colh     0.27\coli    0.44\colj    6.44\colk      2.5E3
\coll      3\colm        \phn7\coln    0.065\eol
\cola    16\colb   51616\colc  204.2799185\cold $-$29.8575701\cole    $-$9.31
\colf     2.43\colg    $-$1.33\colh    $-$0.02\coli     0.20\colj    6.58\colk      1.8E4
\coll      4a\colm        \phn6\coln    0.131\eol
\cola    17\colb   49790\colc  204.2546965\cold $-$29.8585957\cole    $-$7.88
\colf     2.27\colg    $-$0.85\colh     0.27\coli    0.42\colj    6.68\colk      1.1E4
\coll      4a\colm       10\coln    0.091\eol
\cola    18\colb   17035\colc  204.2906967\cold $-$29.8766101\cole    $-$7.23
\colf     2.24\colg    $-$1.48\colh    $-$0.16\coli      0.00\colj     6.70\colk      1.8E3
\coll      4a\colm       15\coln    0.119\eol
\cola    19\colb   66897\colc  204.2654298\cold $-$29.8502774\cole    $-$8.60
\colf     2.96\colg    $-$1.50\colh    $-$0.12\coli      0.00\colj    6.72\colk      4.4E3
\coll      4a\colm       20\coln    0.134\eol
\cola    20\colb   39842\colc  204.2520397\cold $-$29.8651314\cole   $-$10.94
\colf     2.71\colg    $-$1.51\colh    $-$0.15\coli    0.08\colj    6.58\colk      5.2E4
\coll      4a\colm       13\coln    0.086\eol
\cola    21\colb   31457\colc  204.2855284\cold $-$29.8698283\cole    $-$8.83
\colf     2.19\colg    $-$1.31\colh    $-$0.03\coli    0.04\colj    6.78\colk      8.9E3
\coll      4a\colm       17\coln    0.164\eol
\cola    22\colb   74860\colc   204.262148\cold $-$29.8474511\cole    $-$8.59
\colf     2.67\colg    $-$1.43\colh    $-$0.14\coli      0.00\colj    6.78\colk      5.5E3
\coll      4a\colm       21\coln    0.141\eol
\cola    23\colb   89929\colc  204.2581863\cold $-$29.8412261\cole    $-$6.79
\colf     2.23\colg    $-$1.35\colh    $-$0.05\coli    0.04\colj    6.72\colk      1.3E3
\coll      4a\colm       17\coln    0.118\eol
\cola    24\colb   25607\colc  204.2492229\cold $-$29.8725982\cole    $-$7.85
\colf     2.61\colg    $-$0.66\colh     0.42\coli     0.50\colj     6.70\colk      1.2E4
\coll      4a\colm       21\coln    0.120\eol
\cola    25\colb   75041\colc  204.2775285\cold  $-$29.847385\cole    $-$7.31
\colf     2.24\colg    $-$1.61\colh    $-$0.15\coli    0.12\colj    6.52\colk      2.7E3
\coll      4a\colm        \phn7\coln    0.137\eol
\cola    26\colb   64462\colc  204.2811447\cold $-$29.8512236\cole    $-$8.19
\colf     2.30\colg    $-$0.81\colh     0.34\coli    0.46\colj    6.68\colk      1.7E4
\coll      4a\colm        \phn7\coln    0.121\eol
\cola    27\colb   61923\colc   204.263904\cold $-$29.8523354\cole    $-$6.02
\colf     2.52\colg    $-$1.28\colh     0.06\coli    0.36\colj    6.54\colk      1.5E3
\coll      4a\colm        \phn9\coln    0.105\eol
\cola    28\colb   21601\colc  204.2861725\cold $-$29.8744932\cole    $-$9.59
\colf     2.27\colg    $-$1.48\colh    $-$0.12\coli      0.00\colj    6.74\colk      1.5E4
\coll      4a\colm       15\coln    0.158\eol
\cola    29\colb   62771\colc  204.2811659\cold $-$29.8519608\cole    $-$7.25
\colf     2.27\colg    $-$0.96\colh     0.28\coli    0.38\colj     6.70\colk      5.6E3
\coll      4a\colm       16\coln    0.130\eol
\cola    30\colb   13265\colc  204.2547344\cold $-$29.8783208\cole    $-$7.64
\colf     2.44\colg    $-$1.63\colh    $-$0.27\coli      0.00\colj    6.64\colk      2.3E3
\coll      4a\colm       13\coln    0.148\eol
\cola    31\colb    5990\colc  204.2551409\cold $-$29.8813282\cole    $-$8.17
\colf     2.47\colg    $-$1.53\colh    $-$0.26\coli      0.00\colj    6.72\colk      3.9E3
\coll      4a\colm       46\coln    0.124\eol
\cola    32\colb   67958\colc  204.2887307\cold $-$29.8499286\cole    $-$9.84
\colf     3.20\colg    $-$1.35\colh    $-$0.04\coli    0.02\colj    6.78\colk      1.1E4
\coll      4a\colm       19\coln    0.161\eol
\cola    33\colb   66054\colc  204.2644967\cold $-$29.8506016\cole    $-$9.97
\colf     2.63\colg    $-$1.45\colh     0.39\coli      0.00\colj    6.88\colk      2.8E4
\coll      4a\colm       49\coln    0.127\eol
\cola    34\colb   69793\colc  204.2694643\cold $-$29.8493238\cole   $-$11.41
\colf     2.63\colg    $-$1.48\colh    $-$0.16\coli      0.00\colj    6.78\colk      7.5E4
\coll      4a\colm       25\coln    0.136\eol
\cola    35\colb    7716\colc  204.2816067\cold $-$29.8807039\cole    $-$9.17
\colf     2.47\colg    $-$1.43\colh     0.41\coli      0.00\colj    6.88\colk      1.4E4
\coll      4b\colm       68\coln    0.151\eol
\cola    36\colb   36651\colc  204.2524088\cold $-$29.8668398\cole   $-$12.17
\colf     3.09\colg    $-$1.58\colh    $-$0.22\coli      0.00\colj    6.78\colk      1.1E5
\coll      4b\colm      \nodata\rlap{$^e$}\coln    0.070\eol
\cola    37\colb   35937\colc  204.2537197\cold $-$29.8672182\cole   $-$10.31
\colf     2.79\colg    $-$1.24\colh     0.45\coli    0.02\colj     6.90\colk      4.2E4
\coll      4b\colm      \nodata\rlap{$^e$}\coln    0.052\eol
\cola    38\colb   54268\colc  204.2932226\cold $-$29.8563319\cole    $-$8.57
\colf     2.87\colg    $-$0.91\colh     0.79\coli    0.34\colj    6.86\colk      1.7E4
\coll      4b\colm       32\coln    0.108\eol
\cola    39\colb   23366\colc  204.2529648\cold $-$29.8736403\cole    $-$9.39
\colf     2.43\colg    $-$1.29\colh    $-$0.08\coli    0.02\colj    6.78\colk      1.3E4
\coll      4b\colm       41\coln    0.116\eol
\cola    40\colb   27423\colc  204.2896066\cold $-$29.8718058\cole    $-$8.38
\colf     2.34\colg    $-$0.99\colh     0.50\coli    0.36\colj    6.78\colk      1.4E4
\coll      4b\colm       71\coln    0.170\eol
\cola    41\colb   78807\colc  204.2629122\cold $-$29.8459862\cole    $-$9.71
\colf     2.46\colg    $-$1.55\colh    $-$0.23\coli      0.00\colj    6.78\colk      1.6E4
\coll      4b\colm       45\coln    0.169\eol
\cola    42\colb   36729\colc  204.2525635\cold $-$29.8667944\cole   $-$12.15
\colf     3.08\colg    $-$1.45\colh     0.00\coli      0.00\colj    6.74\colk      1.0E5
\coll      4b\colm      \nodata\rlap{$^e$}\coln    0.073\eol
\cola    43\colb   70777\colc  204.2607456\cold $-$29.8489709\cole    $-$8.18
\colf     2.68\colg    $-$1.23\colh     0.49\coli    0.08\colj    6.88\colk      6.7E3
\coll      4b\colm       43\coln    0.087\eol
\cola    44\colb   70769\colc  204.2650559\cold $-$29.8489734\cole    $-$7.77
\colf     2.92\colg    $-$1.34\colh    $-$0.08\coli      0.00\colj    6.76\colk      2.1E3
\coll      4b\colm       18\coln    0.108\eol
\cola    45\colb   37589\colc  204.2519048\cold $-$29.8663287\cole   $-$11.96
\colf     3.20\colg    $-$1.13\colh     0.46\coli    0.24\colj    6.74\colk      1.4E5
\coll      4b\colm      \nodata\rlap{$^e$}\coln    0.064\eol
\cola    46\colb   49893\colc  204.2897929\cold $-$29.8585381\cole    $-$9.76
\colf     2.74\colg    $-$1.49\colh    $-$0.24\coli      0.00\colj    6.78\colk      1.6E4
\coll      4b\colm       48\coln    0.189\eol
\cola    47\colb   37095\colc  204.2523893\cold $-$29.8665896\cole   $-$11.18
\colf     2.79\colg    $-$1.32\colh     0.35\coli      0.00\colj    6.88\colk      8.0E4
\coll      4b\colm      \nodata\rlap{$^e$}\coln    0.059\eol
\cola    48\colb   78154\colc   204.257905\cold $-$29.8462475\cole    $-$8.28
\colf     2.76\colg    $-$0.86\colh     0.18\coli      0.00\colj    7.86\colk      3.4E4
\coll      5a\colm      \nodata\coln    0.080\eol
\cola    49\colb   66216\colc  204.2897363\cold $-$29.8505371\cole    $-$8.87
\colf     2.56\colg    $-$0.98\colh     0.70\coli    0.34\colj    6.84\colk      2.5E4
\coll      5a\colm      \nodata\coln    0.145\eol
\cola    50\colb   76156\colc  204.2556622\cold $-$29.8469956\cole   $-$10.21
\colf     3.15\colg    $-$1.08\colh     0.62\coli    0.06\colj    7.48\colk      1.0E5
\coll      5a\colm      \nodata\coln    0.089\eol
\cola    51\colb   66123\colc  204.2903301\cold $-$29.8505755\cole    $-$9.39
\colf     2.50\colg    $-$1.11\colh     0.59\coli    0.22\colj    6.86\colk      3.0E4
\coll      5a\colm      \nodata\coln    0.125\eol
\cola    52\colb   66069\colc  204.2844254\cold $-$29.8505955\cole    $-$8.88
\colf     3.13\colg    $-$0.78\colh     0.47\coli    0.06\colj    7.81\colk      4.5E4
\coll      5a\colm      \nodata\coln    0.067\eol
\cola    53\colb   85836\colc  204.2694279\cold $-$29.8431197\cole    $-$9.43
\colf     3.25\colg    $-$1.12\colh     0.18\coli     0.10\colj    6.78\colk      8.7E3
\coll      5a\colm      \nodata\coln    0.093\eol
\cola    54\colb   40779\colc  204.2926898\cold $-$29.8645926\cole   $-$10.17
\colf     3.06\colg    $-$0.65\colh     0.49\coli    0.08\colj    7.96\colk      2.1E5
\coll      5a\colm      \nodata\coln    0.044\eol
\cola    55\colb   30950\colc  204.2578713\cold $-$29.8701108\cole   $-$10.85
\colf     2.94\colg    $-$1.08\colh     0.60\coli    0.08\colj    7.49\colk      2.5E5
\coll      5a\colm      \nodata\coln    0.085\eol
\cola    56\colb   83925\colc  204.2643496\cold $-$29.8439131\cole    $-$8.71
\colf     2.79\colg    $-$0.97\colh     0.77\coli    0.14\colj    7.44\colk      4.3E4
\coll      5a\colm      \nodata\coln    0.129\eol
\cola    57\colb   44034\colc  204.2499199\cold $-$29.8626794\cole    $-$9.56
\colf     3.10\colg    $-$0.82\colh     0.44\coli    0.28\colj    6.78\colk      2.1E4
\coll      5b\colm      \nodata\coln    0.030\eol
\cola    58\colb   17159\colc  204.2616672\cold $-$29.8765474\cole    $-$8.22
\colf     3.18\colg    $-$0.42\colh     0.57\coli    0.16\colj    8.06\colk      4.3E4
\coll      5b\colm      \nodata\coln    0.052\eol
\cola    59\colb   25716\colc  204.2827697\cold $-$29.8725452\cole    $-$7.28
\colf     2.76\colg    $-$0.14\colh     0.42\coli      0.00\colj    8.41\colk      2.6E4
\coll      5b\colm      \nodata\coln    0.054\eol
\cola    60\colb   18032\colc  204.2849161\cold  $-$29.876142\cole    $-$7.72
\colf     2.98\colg    $-$0.72\colh     0.61\coli    0.36\colj    6.78\colk      5.7E3
\coll      5b\colm      \nodata\coln    0.064\eol
\cola    61\colb   74692\colc  204.2760088\cold $-$29.8475135\cole    $-$6.99
\colf     3.17\colg    $-$0.04\colh     0.70\coli     0.50\colj    6.74\colk      3.3E3
\coll      5b\colm      \nodata\coln    0.075\eol
\cola    62\colb   85964\colc  204.2565326\cold  $-$29.843069\cole    $-$7.59
\colf     2.99\colg    $-$0.33\colh     0.44\coli      0.00\colj    8.36\colk      2.7E4
\coll      5b\colm      \nodata\coln    0.049\eol
\cola    63\colb   14748\colc  204.2752747\cold $-$29.8776532\cole    $-$7.14
\colf     3.29\colg    $-$0.11\colh     0.40\coli      0.00\colj    8.36\colk      1.2E4
\coll      5b\colm      \nodata\coln    0.038\eol
\cola    64\colb   65733\colc  204.2851579\cold $-$29.8507302\cole    $-$9.82
\colf     2.73\colg    $-$0.94\colh     0.61\coli     0.10\colj    7.59\colk      1.3E5
\coll      5b\colm      \nodata\coln    0.089\eol
\cola    65\colb   65479\colc  204.2862739\cold $-$29.8508316\cole    $-$9.06
\colf     3.19\colg    $-$0.37\colh     0.42\coli      0.00\colj    8.31\colk      7.8E4
\coll      5b\colm      \nodata\coln    0.050\eol
\cola    66\colb   65304\colc  204.2849235\cold $-$29.8508999\cole    $-$8.51
\colf     3.39\colg     0.07\colh     0.80\coli    0.18\colj    8.41\colk      5.8E4
\coll      5b\colm      \nodata\coln    0.039\eol
\cola    67\colb   10114\colc  204.2664853\cold $-$29.8796816\cole    $-$7.17
\colf     2.64\colg    $-$1.53\colh    $-$0.33\coli      0.00\colj    6.64\colk      1.4E3
\coll      5b\colm      \nodata\coln    0.105\eol
\cola    68\colb   58911\colc  204.2807662\cold $-$29.8539005\cole    $-$7.69
\colf     2.88\colg     0.15\colh     1.55\coli    0.32\colj    9.11\colk      2.9E5
\coll      5b\colm      \nodata\coln    0.050\eol
\cola    69\colb   58363\colc  204.2831825\cold $-$29.8541717\cole    $-$7.23
\colf     3.27\colg     0.27\colh     0.54\coli      0.00\colj    8.86\colk      2.9E4
\coll      5b\colm      \nodata\coln    0.034\eol
\cola    70\colb   66553\colc  204.2887134\cold $-$29.8504088\cole    $-$8.06
\colf     3.23\colg     0.64\colh     1.08\coli     0.20\colj    9.01\colk      1.5E5
\coll      5b\colm      \nodata\coln    0.035\eol
\cola    71\colb   55591\colc  204.2710795\cold $-$29.8556304\cole    $-$7.27
\colf     3.14\colg     0.34\colh     0.74\coli    0.04\colj    8.86\colk      4.1E4
\coll      5b\colm      \nodata\coln    0.036\eol
\cola    72\colb   94866\colc  204.2790178\cold $-$29.8386224\cole    $-$8.34
\colf     3.28\colg     0.06\colh     0.39\coli      0.00\colj    8.46\colk      4.0E4
\coll      5b\colm      \nodata\coln    0.031\eol
\cola    73\colb    3113\colc  204.2651882\cold $-$29.8828492\cole    $-$6.93
\colf     3.28\colg     0.04\colh     0.89\coli      0.00\colj    9.16\colk      3.8E4
\coll      6\colm      \nodata\coln    0.031\eol
\cola    74\colb   12505\colc  204.2762954\cold $-$29.8786591\cole    $-$6.67
\colf     3.26\colg     0.44\colh     0.97\coli      0.00\colj    9.06\colk      2.2E4
\coll      6\colm      \nodata\coln    0.029\eol
\cola    75\colb   24528\colc  204.2890323\cold $-$29.8730866\cole    $-$9.15
\colf     3.13\colg    $-$0.14\colh     0.80\coli    0.22\colj    8.36\colk      1.8E5
\coll      6\colm      \nodata\coln    0.040\eol
\cola    76\colb   55985\colc  204.2695705\cold $-$29.8553997\cole    $-$7.97
\colf     2.91\colg     0.02\colh     1.01\coli      0.00\colj    9.26\colk      2.1E5
\coll      6\colm      \nodata\coln    0.040\eol
\cola    77\colb   18044\colc  204.2956504\cold $-$29.8761376\cole    $-$8.29
\colf     3.24\colg     0.12\colh     0.54\coli      0.00\colj    8.61\colk      5.3E4
\coll      6\colm      \nodata\coln    0.034\eol
\cola    78\colb   54416\colc  204.2898224\cold $-$29.8562536\cole    $-$8.01
\colf     3.15\colg    $-$0.37\colh     0.91\coli    0.38\colj    7.96\colk      6.1E4
\coll      6\colm      \nodata\coln    0.042\eol
\cola    79\colb   46572\colc  204.2634717\cold $-$29.8608741\cole    $-$8.54
\colf     3.05\colg     0.18\colh     0.81\coli    0.04\colj    8.91\colk      1.7E5
\coll      6\colm      \nodata \coln    0.058\eol

\enddata
\tablecomments{
$^a$ Values of M$_v$ throughout this paper assume a distance modulous m -
M = 28.28, external extinction Av = 0.229, no correction for internal
extinction, and size-dependent aperture corrections described in
Chandar et~al.\ (2010).   Only external exctinction corrections have been made for values of $U\!-\!B$ and $V\!-\!I$.\\
$^b$ Concentration Index, defined as the magnitude difference between 0.5 and
3~pix radii apertures. \\
$^c$   Morphological category as defined in Section~2.\\
$^d$ Surface brightness fluctuations measured using the technique described in Section~4. A 4~pixel radius was used for objects 66, 69, 70, 79 rather than the normal 10 pixel radius due to the presence of a likely unrelated bright star within a 10 pixel radius.     \\
$^e$ The five category 4b clusters with no measured values of H$\alpha$
are all in the large bubble-like structure below the nucleus of M83 (see Figure~2 in Chandar et~al.) 2010. No estimates of R(H$\alpha$) are provided for these clusters since the bubble appears to be formed by the integrated effects of a large number of clusters rather than any one individual cluster.   \\
}
\end{deluxetable}

\clearpage
\begin{deluxetable}{cccccccc}
\tabletypesize{\small}
\rotate
\tablecaption{Mean Values for Morphological Categories \label{tab:table_2}}
\tablewidth{0pt}
\tablehead{
\colhead{Category} & \colhead{Mean M$_v$} & \colhead{Mean CI}  & \colhead{Mean $E(B\!-\!V$)} & \colhead{Mean  Log Age} & \colhead{Mean Mass} & \colhead{Mean R(H$\alpha$)} & \colhead{Mean RMS} \\
 & (mag) & (mag) & (mag) & (yr) & M$\odot$ & (pc) }
\startdata
3\phn &        $-7.29\pm1.30$ &  $2.56\pm0.15$ &  $0.42\pm0.10$ &  $6.53\pm0.08$ &  $1.5E4\pm3.0E4$ &    $\phn5.0\pm2.3$\phn &  $0.088\pm0.021$      \\
4a &       $-8.24\pm1.29$ &  $2.43\pm0.24$ &  $0.17\pm0.19$ &  $6.67\pm0.09$ &  $1.2E4\pm1.4 E4$ &  $13.7\pm5.4$\phn &  $0.125\pm0.023$      \\
4b &       $-9.63\pm1.58$ &  $2.71\pm0.29$ &  $0.08\pm0.14$ &  $6.79\pm0.07$ &  $3.6E4\pm4.2E4$ &  $38.1\pm18.5$ &  $0.118\pm0.040$      \\
5a &        $-9.42\pm0.84$ &  $2.90\pm0.20$ &  $0.12\pm0.10$ &  $7.39\pm0.46$ &  $8.3E4\pm8.8E4$ &  \nodata  &  $0.095\pm0.032$      \\
5b &       $-7.98\pm0.88$ &  $3.08\pm0.22$ &  $0.13\pm0.16$ &  $8.05\pm0.86$ &  $6.0E4\pm7.5E4$ & \nodata &  $0.052\pm0.022$      \\
6\phn &        $-7.94\pm0.87$ &  $3.15\pm0.13$ &  $0.09\pm0.15$ &  $8.76\pm0.47$ &  $\llap{1}0.4E4\pm7.8E4$ &  \nodata &  $0.039\pm0.010$      \\

\enddata

\end{deluxetable}

\clearpage
\begin{deluxetable}{rrcccccrc}
\newcommand\cola {\null}
\newcommand\colb {&}
\newcommand\colc {&}
\newcommand\cold {&}
\newcommand\cole {&}
\newcommand\colf {&}
\newcommand\colg {&}
\newcommand\colh {&}
\newcommand\coli {&}
\newcommand\colj {&}
\newcommand\eol{\\}
\newcommand\extline{&&&&&&&&&&&&&\eol}
\tablecaption{Parameters for 22 ``Single Star'' HII Regions \label{tab:table_3}}
\tablewidth{0pt}
\tablehead{
\colhead{\#} & \colhead{ID} & \colhead{RA} & \colhead{Decl.} & \colhead{M$_V$} & \colhead{CI} & \colhead{$U\!-\!B$} & \colhead{$V\!-\!I$} & \colhead{R(H$\alpha$)} \\
& & & & (mag) & (mag) & (mag) & (mag) & (pc)}

\startdata

\cola     1\colb    3550\colc  204.2548844\cold $-$29.8825632\cole    $-$5.51
\colf     2.13\colg    $-$1.63\colh    $-$0.22\coli    9\eol
\cola     2\colb    3916\colc  204.2554865\cold $-$29.8823689\cole    $-$5.19
\colf     2.30\colg    $-$1.60\colh    $-$0.28\coli    8\eol
\cola     3\colb   10373\colc  204.2549063\cold $-$29.8795771\cole    $-$5.24
\colf     2.28\colg    $-$1.49\colh    $-$0.16\coli    7\eol
\cola     4\colb   11098\colc   204.2672170\cold $-$29.8792752\cole    $-$6.89
\colf     2.02\colg    $-$1.64\colh    $-$0.33\coli   \llap{1}4\eol
\cola     5\colb   29250\colc  204.2745636\cold $-$29.8709998\cole    $-$4.41
\colf     2.09\colg    $-$1.21\colh     0.02\coli    8\eol
\cola     6\colb   35101\colc  204.2745919\cold $-$29.8676963\cole    $-$5.23
\colf     2.11\colg    $-$1.49\colh    $-$0.39\coli    6\eol
\cola     7\colb   37778\colc  204.2667959\cold $-$29.8662227\cole    $-$5.93
\colf     2.21\colg    $-$1.13\colh     0.08\coli    9\eol
\cola     8\colb   47544\colc  204.2823703\cold $-$29.8602099\cole    $-$5.86
\colf     2.10\colg    $-$1.51\colh    $-$0.14\coli    7\eol
\cola     9\colb   47772\colc  204.2689918\cold  $-$29.8600630\cole    $-$5.09
\colf     2.12\colg    $-$1.31\colh     0.27\coli    7\eol
\cola    10\colb   48660\colc  204.2894733\cold $-$29.8594131\cole    $-$4.26
\colf     2.25\colg    $-$1.62\colh    $-$0.48\coli    2\eol
\cola    11\colb   49679\colc  204.2535423\cold $-$29.8586761\cole    $-$5.78
\colf     2.18\colg    $-$0.49\colh     0.32\coli    9\eol
\cola    12\colb   50923\colc  204.2552793\cold $-$29.8579399\cole    $-$6.13
\colf     2.23\colg    $-$1.15\colh     0.25\coli    5\eol
\cola    13\colb   51301\colc  204.2887045\cold $-$29.8577311\cole    $-$5.19
\colf     2.18\colg    $-$1.27\colh     0.15\coli    2\eol
\cola    14\colb   52904\colc  204.2462139\cold $-$29.8569822\cole    $-$4.69
\colf     2.23\colg    $-$1.59\colh     0.14\coli    1\eol
\cola    15\colb   56031\colc  204.2828693\cold $-$29.8553764\cole    $-$6.15
\colf     2.18\colg    $-$1.36\colh     0.02\coli    6\eol
\cola    16\colb   56485\colc  204.2849019\cold $-$29.8551313\cole    $-$6.19
\colf     2.04\colg    $-$1.42\colh     0.06\coli    8\eol
\cola    17\colb   56492\colc  204.2829518\cold $-$29.8551272\cole    $-$6.24
\colf     2.24\colg    $-$0.12\colh     1.07\coli    9\eol
\cola    18\colb   56937\colc  204.2841803\cold $-$29.8548844\cole    $-$8.06
\colf     2.13\colg    $-$0.21\colh     0.90\coli    6\eol
\cola    19\colb   60624\colc  204.2516243\cold $-$29.8530186\cole    $-$4.99
\colf     2.20\colg    $-$1.28\colh    $-$0.02\coli    7\eol
\cola    20\colb   61345\colc  204.2496746\cold $-$29.8526196\cole    $-$4.79
\colf     2.23\colg    $-$1.25\colh    $-$0.32\coli    4\eol
\cola    21\colb   72114\colc   204.2810230\cold $-$29.8485123\cole    $-$6.34
\colf     2.08\colg    $-$0.77\colh     0.33\coli   \llap{1}2\eol
\cola    22\colb   88339\colc  204.2804302\cold $-$29.8420518\cole    $-$4.49
\colf     2.30\colg    $-$1.08\colh    $-$0.13\coli    7\eol

\enddata
\end{deluxetable}

\end{document}